\documentclass[twocolumn]{aa} 

\newcommand{\xpos}{\ensuremath{x_0}}
\newcommand{\ypos}{\ensuremath{y_0}}

\newcommand{\VROT}{{\sc vrot}}
\newcommand{\nodata}{ ~$\cdots$~ }
\def\vvf{\ensuremath{V_{\rm vf}}}
\def\vpv{\ensuremath{V_{\rm pv}}}
\def\ML{\ensuremath{\Upsilon_{*}}}
\newcommand{\kms}{\ensuremath{\,\mbox{km}\,\mbox{s}^{-1}}}
\newcommand{\Msun}{$M_{\odot}$}
\newcommand{\HI}{H\,{\sc i}}
\newcommand{\ihi}{\ensuremath{i_{\rm H{\sc I}}}}
\newcommand{\ikin}{\ensuremath{i_{\mathrm{kin}}}}
\newcommand{\iopt}{\ensuremath{i_{\mathrm{opt}}}}
\newcommand{\rotcur}{{\sc rotcur}}

\def\20{\ensuremath{W_{20}}}
\def\50{\ensuremath{W_{50}}}

\usepackage{graphicx}
\usepackage{natbib}
\usepackage{booktabs}
\usepackage{morefloats,epsf,amsmath,subfigure}
\bibpunct{(}{)}{;}{a}{}{,}
\usepackage{textcomp}
\usepackage{multirow}
\usepackage{txfonts}
\begin{document}

\titlerunning{Baryonic Tully-Fisher relation}

   \title{The baryonic Tully-Fisher relation and its implication for dark matter halos}

   \author{C. Trachternach
          \inst{1}
          \and
          W.J.G. de Blok\inst{2}
                    \and
          S.S. McGaugh\inst{3}
      \and
         J.M. van der Hulst\inst{4}
	\and
          R.-J. Dettmar\inst{1}
          }

   \offprints{W.J.G. de Blok}

   \institute{Astronomisches Institut, Ruhr-Universit\"at
  Bochum, Universit\"atsstra{\ss}e 150, 44780 Bochum, Germany  
         \and
             Department of Astronomy, University of Cape Town, Private Bag
X3, Rondebosch 7701, South Africa
            \and
Department of Astronomy, University of Maryland, College Park, MD 20742-2421
            \and
Kapteyn Astronomical Institute, University of Groningen, Postbus 800, 9700 AV Groningen, The Netherlands
}

   \date{}

   \abstract {The baryonic Tully-Fisher relation (BTF) is a
     fundamental relation between baryonic mass and maximum rotation
     velocity. It can be used to estimate distances, as well as to
     constrain the properties of dark matter and its relation with the
     visible matter.}
   {In this paper, we explore if extremely low-mass dwarf galaxies
     follow the same BTF relation as high-mass galaxies. 
     We quantify the scatter in the BTF relation and use this to
     constrain the allowed elongations of dark matter halo potentials.}
   {We obtained \HI\ synthesis data of 11 dwarf galaxies and derive
     several independent estimates for the maximum rotation velocity.}
   {Constructing a BTF relation using data from the literature for the
     high-mass end, and galaxies with detected rotation from our
     sample for the low-mass end results in a BTF with a scatter of
     0.33 mag.}
   {This scatter constrains the ellipticities of the potentials in the
     plane of the disks of the galaxies to an upper limit of 0-0.06,
     indicating that dwarf galaxies are at most only mildly
     tri-axial. Our results indicate that the BTF relation is a
     fundamental relation which all rotationally dominated galaxies
     seem to follow.  }


   \maketitle

%

\section{Introduction}
The Tully-Fisher (TF) relation \citep{tully-1977}, the relation
between the luminosity and rotation velocity of a galaxy, has been
extensively used to estimate extragalactic distances
\citep[e.g.,][]{pierce-1988,willick-1997,sakai-2000,tully-2000,springob-2007}.
In the usual ``classical'' interpretation, luminosity is a proxy for
the stellar mass, which in turn depends on the total (visible and
dark) mass and through it on the rotation velocity.  The slope and
zero point of this classical TF relation do not depend on the central
surface brightness of galaxies \citep{zwaan-1995}, though for very low
mass dwarf galaxies, the slope tends to steepen
\citep[e.g.,][]{matthews-1998b, mcgaugh-2000}. Low-mass dwarf galaxies
are apparently underluminous given their rotation velocity and
therefore fall below the TF relation as defined by the high mass
galaxies. A single linear relation can be restored if one replaces the
luminosity (or stellar mass) with the baryonic disk mass, thus
including the gas mass as well \citep{mcgaugh-1999,
  mcgaugh-2000}. This relation is called the baryonic Tully-Fisher
(BTF) relation and has been studied by many authors in the last few
years \citep[e.g.,][]{bell-2001, verheijen-2001, gurovich-2004,
  mcgaugh-2004,mcgaugh-2005, geha-2006, derijcke-2007,
  noordermeer-2007,stark-2009}.

The existence of a (baryonic) Tully-Fisher relation places severe
constraints on galaxy formation and evolution theories
\citep[cf.][]{eisenstein-1996, mcgaugh-1998a, mcgaugh-1998b, mo-1998,
  steinmetz-1999, blanton-2007}.  For example, \cite{franx-1992} note
that the observed scatter in the TF relation places upper limits on
the elongation of dark matter halos.  They find that that the
ellipticity of the potential in the plane of the disk is most likely
between 0 and $\sim 0.06$. This is in good agreement with what
\cite{trachternach-08} found observationally for a sample of 18 dwarf
and spiral galaxies from the THINGS survey \citep{walter-07,
  deblok-07}, but in disagreement with results from CDM simulations
which predict more elongated potentials \citep[e.g.,][]{frenk-1988,
  hayashi-2006}.  The smaller scatter in the observed BTF relation
\citep[e.g.,][]{mcgaugh-2005}, means that it can be used to put
similar constraints on lower mass galaxies as well.  This was recently
shown by \citet{stark-2009} who studied the BTF for a large sample of
dwarf galaxies with resolved \HI\ rotation curves, and found a linear
BTF relation with small scatter.

The low mass end of the BTF relation has also been studied by
\cite{geha-2006} and \cite{kovac-2007}.  They measure line widths and
after correcting these for broadening due to turbulent motion of the
\HI, they find that the extreme dwarf galaxies of their sample follow
the same BTF relation as the high mass galaxies, albeit with a larger
scatter. This increased scatter is most likely a result of their use
of the \20\ profile.  \cite{broeils-1992} and \cite{verheijen-1997}
already pointed out that using the maximum rotation velocity from a
resolved rotation curve significantly decreases the scatter as
compared to using line width measurements. 

In this paper, we attempt to determine several independent estimates
for the maximum rotation velocity ($V_{\mathrm{max}}$) for a sample of
extremely low-mass dwarf galaxies.  Doing this, we can check whether
these galaxies follow the same tight correlation between baryonic disk
mass and rotation velocity as their high mass counterparts.  As a
reference, we use the sample of \cite{mcgaugh-2005}, for which
well-determined estimates for $V_{\mathrm{max}}$ exist based on
analysis of well-resolved rotation curves. For an in-depth description
of the comparison sample, the reader is referred to
\cite{mcgaugh-2005} and references therein.

The paper is organized as follows: in Section 2 we describe the
observations and the data reduction, which is followed by a
description of the different methods in estimating $V_{\mathrm{max}}$
in Section 3.  We comment the individual galaxies in Section 4, and
present further analysis and our results in Section 5. We summarize
our results and give our conclusions in Section 6. The Appendix
contains moment maps, position-velocity diagrams and channel maps of
our sample galaxies.

\begin{table*}
\begin{minipage}[t]{\linewidth}
\caption[Mapping/Noise parameters]{Properties of the data.}
\label{table:btf:mapping-params}
\centering
\small
\renewcommand{\footnoterule}{}  
\begin{tabular}{lcrrrrcccc}
  \toprule
  \multicolumn{1}{c}{ID} & Date & \multicolumn{1}{c}{length} & \multicolumn{1}{c}{${b_{\rm maj}}$} &  \multicolumn{1}{c}{${b_{\rm min}}$}  &\multicolumn{1}{c}{PA} & noise/channel & pixel size & channel width \\
  ~ & of Obs. & \multicolumn{1}{c}{of Obs.} & \multicolumn{1}{c}{\arcsec} & \multicolumn{1}{c}{\arcsec} & \multicolumn{1}{c}{\degr}  & ${\rm mJy\ beam^{-1}}$ & \arcsec & \kms\\
  \multicolumn{1}{c}{(1)} & (2) & \multicolumn{1}{c}{(3)} & \multicolumn{1}{c}{(4)} & \multicolumn{1}{c}{(5)} & \multicolumn{1}{c}{(6)} & (7) & (8) & (9) \\
  \midrule
  D500-2  & 01-05-2002 & 12h & 25.0 & 10.6 & 0.0 & 1.0 & 5 & 4.12 \\
  D500-3  & 03-05-2002 & 12h & 33.0 & 12.3 & 0.0 & 0.6 & 5 & 4.12  \\
  D512-2  & 16-05-2004 & 12h & 28.9 & 13.4 & 0.6 & 0.9 & 4 & 2.10 \\
  D564-8  & 08-05-2002 & 12h & 36.4 & 12.8 & 0.0 & 0.6 & 5 & 4.12\\
  D572-5  & 10-05-2004 & 12h & 86.6 & 24.0 & $-$0.2 & 1.2 & 4 & 2.10 \\
  D575-1\footnote{The data for D575-1 are combined from two individual observations} & 07-05-2004 & 12h & 48.2 & 18.4 & 0.2 & 0.9 & 4 & 2.10 \\
  \nodata   & 12-05-2004 & 12h &\nodata & \nodata&\nodata &\nodata & \nodata& \nodata&\\
  D575-2\footnote{The data for D575-2 are combined from three individual observations} & 09-05-2004 & 12h & 45.6 & 14.0 & $-$3.5 & 0.6 & 4 & 2.10\\
  \nodata& 13-05-2004 & 10h &\nodata &\nodata &\nodata &\nodata &\nodata &\nodata &\\
  \nodata& 17-10-2004 & ~7h & \nodata&\nodata &\nodata &\nodata &\nodata &\nodata &\\
  D575-5  & 05-05-2004 & 12h & 37.5 & 11.3 &$-$0.1 & 1.1 & 4 & 2.10 \\
  D631-7  & 04-11-2004 & 12h & 57.7 & 12.0 & 0.2 & 1.1 & 4 & 2.10 \\
  D640-13 & 10-11-2004 & 12h & 70.8 & 11.5 & $-$0.1 & 1.1 & 4 & 2.10 \\
  D646-7  & 11-05-2004 & 12h & 133.8 & 20.6 & 0.6 & 1.4 & 4 & 2.10 \\
  \bottomrule
\end{tabular}
\end{minipage} 

{\sc Notes:} (1): galaxy identifier; (2): date of
observations; (3): length of observations; (4, 5) major and minor axis
diameter of the robust weighted beam in arcsec; (6): position angle of
the beam (in degrees), measured counter-clockwise from the north; (7):
noise per channel in mJy beam$^{-1}$; (8): pixel size in arcsec; (9):
channel width in \kms
\end{table*}

\section{The data}\label{btf:sec:data}
Our sample was selected from the larger sample of
\cite{schombert-1997}, which is one of the largest samples of extreme
field dwarf galaxies for which both line width measurements and \HI\
masses \citep{eder-2000}, as well as optical photometry
\citep{pildis-1997} exist. The galaxies were chosen to be relatively
nearby ($v_{\mathrm{hel}}< 1400\kms$), have suitable optical
inclinations for potential derivation of their rotation curves ($45 \le \iopt
\le 75$), and to have $V$- and $I$-band photometry available.

\subsection{Observations}
Observations were carried out at the Westerbork Synthesis Radio
Telescope (WSRT) in maxi-short configuration, in the 21-cm line of
neutral hydrogen. We use both polarizations and sample 1024 channels
with a bandwidth of 10 MHz or 20 MHz (corresponding to velocity
resolutions of 2.10 or 4.12 \kms). The sample integration time was set
to 60 seconds. Further observational details are summarized in
Table~\ref{table:btf:mapping-params}.

\subsection{Data reduction}
The calibration and data reduction of the data is performed using
standard routines in MIRIAD\footnote{Multichannel Image
  Reconstruction, Image Analysis and Display \citep{sault-1995}}.  The
data are calibrated using one of the standard primary calibrators used
at the WSRT (3C48, 3C147, 3C286, J2052+362). The primary calibrators
are also used for the bandpass and gain corrections.

The line-free channels (i.e., the channels containing only continuum
emission) were used to create a continuum image. This image was used
to self-calibrate the data.  After the self-calibration and continuum
subtraction, image cubes were created using the robust weighting
scheme \citep{briggs-1995} with a robust parameter of zero for all
galaxies. These image cubes were then cleaned down to a level of
$\sim$\,1$\sigma$. For a list of the beam sizes and noise levels, see
Table~\ref{table:btf:mapping-params}.  We additionally created Hanning
smoothed data cubes, which were used for the creation of the \HI\
profiles and the derivation of the velocity widths.

Using GIPSY\footnote{GIPSY, the Groningen Image Processing SYstem
  \citep{vanderhulst-1992}}, we created zeroth, first and second
moment maps of all galaxies.  In order to isolate significant signal,
we smoothed the data cubes to half the original resolution and only
retained pixels with values $>2.5$ times the (smoothed) noise
value. Spurious pixels were blotted by hand. This smoothed and blotted
data cube was used as a mask for the original data cube. We determined
the number of channels with significant emission contributing to each
unblanked pixel in the unsmoothed moment maps and created a map
containing the signal-to-noise (S/N) of each pixel in the total
intensity (zeroth moment) map.  Using this map, we determined the
average pixel value in the zeroth moment map corresponding
to a $\mathrm{S/N}=3$, and clip all moment maps using this flux limit.
Channel maps and moment maps are presented in the Appendix.

\section{Estimating the maximum rotation velocity}\label{btf:sec:Vmax-estimates}
In order to construct a (baryonic) Tully-Fisher relation, one needs to
estimate the maximum (outer) rotation velocity ($\rm{V_{max}}$) of a
galaxy.  There are several ways in which this can be done, which are
described below in increasing order of preference.

\subsection{\HI\ velocity profile}
The simplest way is to use the width $W$ of the global \HI\ velocity
profile, usually measured at the 20 (50) percent level of the maximum
intensity, and denoted as \20\ (\50).  The advantage in using the
profile width is that it is easy to measure and can be derived using
low resolution data. The drawback is that one cannot distinguish
between rotation and turbulence.  This uncertainty matters little for
large, fast rotating spiral galaxies. However, as one goes to smaller
and more slowly rotating galaxies, turbulent motions will start to
contribute significantly to the total width of \HI\ profiles
\citep[cf.][]{verheijen-1997}. In our analysis, we use the velocity
profiles from the Hanning smoothed data cubes and correct them for
instrumental broadening and turbulent motion. The corrections applied
are addressed and discussed more fully in
Section~\ref{sec:btf:lw-corrections}.

\subsection{Major axis position-velocity diagram}
An alternative way to derive ${V_{\rm max}}$ is to make use of the
major axis position-velocity ($pV$) diagram. For a galaxy with a flat
rotation curve, Gaussian fits to the outer galaxy radii in the major
axis \emph{pV}-diagram (see, e.g., Fig.~\ref{fig:btf:d500-2} in the
Appendix) yield an estimate for the (projected) amplitude of the
rotation.  The maximum rotation velocity is then calculated as half
the difference in velocity between approaching and receding sides,
corrected for inclination:
\begin{equation}
V_{pv}=\frac{V_{\rm{receding}} - V_{\rm{approaching}}}{2\,\sin(i)}
\label{eq:btf:V-pv}
\end{equation}
The \emph{pV}-diagram is potentially able to better constrain the
the maximum rotation curve velocity than  \20\ and \50\ values
\citep{verheijen-2001}.

\subsection{Tilted-ring models}

A third method for the derivation of $V_{\rm max}$ is a tilted-ring
model, in which the kinematics of a galaxy are described using a set
of concentric rings. Each of these rings can have its own center
position $(x_0, y_0)$, systemic velocity $V_{\rm sys}$, rotation
velocity $V_{\rm rot}$, inclination $i$, and position angle PA.  The
pre-requisite for this method is a resolved velocity field showing
signs of rotation.

For the galaxies in our sample that meet this criterion, we derive
rotation curves using the GIPSY task \rotcur.  Assuming that the gas
moves on circular orbits, the line-of-sight velocity can be expressed
as:
\begin{equation}
 V(x,y)=V_{\rm sys}+V_{\rm rot}(r)\sin(i)\cos(\theta).
\label{eq:btf:Vlos}
\end{equation}
Here, $\theta$ is the azimuthal distance from the major axis in the
plane of the galaxy and is related to the position angle PA of the
galaxy as measured in the plane of the sky by
\begin{equation}
 \cos(\theta)=\frac{-(x-x_0)\sin(PA)+(y-y_0)\cos(PA)}{r}
\end{equation}
and
\begin{equation}
\sin(\theta)=\frac{-(x-x_0)\cos(PA)-(y-y_0)\sin(PA)}{r\cos(i)}.
\end{equation}
The PA is measured counter-clockwise from the north to the major axis
of the receding side of the galaxy.
As positions along the major axis of a galaxy carry more rotational
information than positions near the minor axis, we weight the individual
data points by $|\cos(\theta)|$.

The derivation of a rotation curve is generally an iterative process
involving the consecutive determination of the various tilted-ring
parameters. Following is a general description of the procedure
applied.  In order to get good initial estimates for $i$, PA, and the
center position, we fit isophotes at varying intensity levels to the
\HI\ total intensity maps, taking care that the results are not
affected by small-scale structures.  As initial estimates for $V_{\rm
  sys}$ and $V_{\rm rot}$, we use the central velocities of the \50\
profile and $\frac{1}{2}$\50$/\sin(i)$, respectively.

In the first tilted-ring fit, we determine the systemic velocity by
keeping all parameters except $V_{\rm sys}$ and $V_{\rm rot}$ fixed.
In a second run, we derive the position of the dynamical center
leaving by leaving (only) the central position of the rings and
$V_{\rm rot}$ unconstrained.  In a third fit, we determine the
position angle by leaving only PA, $i$ and $V_{\rm rot}$ as free
parameters. As in most cases we sample the rotation curves with
relatively few ($\sim$\,10) tilted-rings, we approximate the fitted
values either by a constant or linearly changing PA.  Once the PA is
modeled in such a way, we make another fit with only $i$ (and $V_{\rm
  rot}$) as free parameters and derive the inclination of the
galaxy. We do not model any radial trends for the inclination. In a
last run, we determine the rotation curve by keeping all parameters
except $V_{\rm rot}$ fixed at their best determined values.  The
maximum velocity of the rotation curve derived in this way will be
referred to as \vvf\ hereafter.

\begin{table*}
\begin{minipage}[t]{\linewidth}
\caption[``Profile-width'' sub-sample: derived parameters]{Derived parameters for the profile width sub-sample}
\label{table:btf:derived-params}
\centering
\scriptsize
\renewcommand{\footnoterule}{}  
\begin{tabular}{lcccrcccccccl}
\toprule
\multicolumn{1}{c}{ID} & $\alpha_{2000}$ & $\delta_{2000}$ & ${V_{\rm sys}}$ & \multicolumn{1}{c}{D} & $M_V$ & $(V-I)$ & $M_{HI}$ & \20 & \50 & \20$_{\rm{,~turb}}$ & \50$_{\rm{,~turb}}$ & \multicolumn{1}{c}{$i$}\\
~ & (h m s) & ($\circ\,\,\, \arcmin\,\,\, \arcsec$) & \kms & \multicolumn{1}{c}{Mpc} & mag  & & $10^7$ \Msun & \kms & \kms & \kms & \kms & \multicolumn{1}{c}{\degr}\\
\multicolumn{1}{c}{(1)} & (2) & (3) & (4) & \multicolumn{1}{c}{(5)} & (6) & (7) & (8) & (9) & (10) & (11) & (12) & (13)\\

\midrule
D572-5  & 11 48 16.4 & +18 38 33 & 994  & 14.6 & $-$14.56 & 0.52 & 8.55 & 77 & 62 & 68 & 55 & 50\footnote{optical \emph{I}-band inclination from \cite{pildis-1997}}\\ 
D575-1  & 12 51 46.1 & +21 44 07 & 600  & 10.0 & $-$14.22 & 0.70 & 3.76 & 38 & 24 & 30 & 18 & 53\\ 
D575-5  & 12 55 41.4 & +19 12 34 & 437  & 7.7 & $-$13.21 & 0.44 & 4.27 & 28 & 19 & 16 & 10 & 50\\ 
D640-13 & 10 56 13.9 & +12 00 41 & 958  & 13.4 & $-$14.36 & 0.55 & 4.53 & 38 & 25 & 30 & 18 & 48$^a$\\ 
D646-7  & 12 58 40.4 & +14 13 03 & 233  & 2.1\footnote{distance as given in \cite{karachentsev-2003}} & $-$12.52 & 0.84 & 0.37 & 35 & 23 & 26 & 16 & 55$^a$\\ 
\bottomrule
\end{tabular}
\end{minipage} {\sc Notes:} (1): galaxy identifier; (2, 3): central
position taken from NED; (4): systemic velocity derived from the
central velocity of the \50\ profile; (5): distance in Mpc. If no
reference is given, the distance is based on $V_{\mathrm{sys}}$
(column 4) and a Hubble flow using $H_0=75\mathrm{\kms Mpc^{-1}}$,
including a correction for Virgocentric infall \citep{mould-2000};
(6): absolute \emph{V}-band magnitude as given in \cite{pildis-1997},
corrected to our distance estimates; (7): $V$-$I$ color from
\cite{pildis-1997}; (8): total \HI\ mass (in units of $10^7
M_{\sun}$); (9): uncorrected width of the \HI\ profile at the 20
percent level of the maximum intensity. The values listed here for
\20\ are derived using the Hanning smoothed data; (10): as
column 9, but for the \50\ profile; (11): width of the \20 profile,
corrected for the finite velocity resolution, and turbulent motion of
the \HI\ gas; (12): as column 11, but for the \50\ profile; (13):
adopted inclination angle. The velocity widths in Cols. (9) - (12) are
not corrected for inclination effects.
\end{table*}

\begin{table*}
\begin{minipage}[t]{\linewidth}
  \caption[``Rotation curve'' sub-sample: derived parameters]{Derived
    parameters for the rotation curve sub-sample}
\label{table:btf:rotcur-params}
\centering
\scriptsize

\renewcommand{\footnoterule}{} 
\begin{tabular}{lccrrccrrrrrlrcc}
\toprule

\multicolumn{1}{c}{ID}& $\alpha_{2000}$ & $\delta_{2000}$ & \multicolumn{1}{c}{${V_{\rm sys}}$} & \multicolumn{1}{c}{D} & $M_V$ & $(V-I)$ & \multicolumn{1}{c}{$M_{HI}$} & \multicolumn{1}{c}{\20} & \multicolumn{1}{c}{\50} &\multicolumn{1}{c}{ \20$_{\rm{,~turb}}$} & \multicolumn{1}{c}{\50$_{\rm{,~turb}}$} & \multicolumn{1}{c}{$i$} & \multicolumn{1}{c}{$\langle PA \rangle$} & \vvf & \vpv\\
~ & (h m s) & ($\circ\,\,\, \arcmin\,\,\, \arcsec$) & \multicolumn{1}{c}{\kms} & \multicolumn{1}{c}{Mpc} & mag  & & \multicolumn{1}{c}{$10^7$ \Msun} & \multicolumn{1}{c}{\kms} & \multicolumn{1}{c}{\kms} & \multicolumn{1}{c}{\kms} & \multicolumn{1}{c}{\kms} &\multicolumn{1}{c}{\degr} &\multicolumn{1}{c}{\degr} & \kms & \kms\\
\multicolumn{1}{c}{(1)} & (2) & (3) & \multicolumn{1}{c}{(4)} & \multicolumn{1}{c}{(5)} & (6) & (7) &\multicolumn{1}{c}{(8)} & \multicolumn{1}{c}{(9)} & \multicolumn{1}{c}{(10)} & \multicolumn{1}{c}{(11)} & \multicolumn{1}{c}{(12)} & \multicolumn{1}{c}{(13)} & \multicolumn{1}{c}{(14)} & (15) & (16)\\
\midrule
D500-2$^*$ & 10 31 43.0 & +25 18 33 & 1259 & 17.9~ & $-$16.38 & 0.42 & 87.08 & 135 & 120 & 119 & 105 & 57 $\pm$ 6 & 345 & 68 & 68\\ 
D500-3  & 10 05 59.5 & +23 52 04 & 1327 & 18.5~ & $-$15.74 & 0.31 & 6.58 & 96 & 83 & 84 & 72 & 55\footnote{optical \emph{I}-band inclination from \cite{pildis-1997}} $\pm$ 6 & \nodata & \nodata & 45\\ 
D512-2 & 14 33 20.2 & +26 59 54 & 840  & 14.1~ & $-$15.22 & 0.80 & 6.96 & 88 & 74 & 78 & 65 & 56 $\pm$ 10 & 40 & 35 & 37\\ 
D564-8$^*$ & 09 02 54.0 & +20 04 28 & 478  & 6.5~ & $-$12.64 & 0.93 & 1.58 & 62 & 48 & 54 & 42 & 63 $\pm$ 7 & 13 & 25 & 29\\ 
D575-2$^*$ & 12 52 21.9 & +21 37 46 & 774  & 12.2~ & $-$15.15 & 0.78 & 31.51 & 146 & 119 & 129 & 104 & 57 $\pm$ 5 & 220 & 74 & \nodata\\ 
D631-7$^*$ & 07 57 01.8 & +14 23 27 & 311  & 5.5\footnote{distance as given in \cite{karachentsev-2003}} & $-$14.50 & 0.55 & 14.68 & 113 & 89 & 100 & 78 & 59 $\pm$ 3 & 324 & 58 & 53\\ 
\bottomrule
\end{tabular}
\end{minipage} {\sc Notes:} Cols. (1) to (12) as in
Table~\ref{table:btf:derived-params}, except that for
the starred galaxies, the center position and $V_{\mathrm{sys}}$ were
derived kinematically; (13) inclination angle (optical inclination for
D500-3, kinematic inclination otherwise). The uncertainties given for
the kinematic inclination angles represent the scatter of the values
of the individual tilted-rings. For D500-3, 
the average uncertainty of the kinematic inclination estimates is given; (14)
average position angle of the rotation curve analysis, measured
counter-clockwise from north to the receding side of the galaxy; (15)
maximum rotation velocity from the tilted-ring analysis of the
velocity field, corrected for inclination; (16) maximum
rotation velocity from the position-velocity diagram, corrected for
inclination.
\end{table*}

\section{Comments on individual galaxies}\label{btf:sec:indiv-galaxies}
In this section, we present our results for the individual galaxies.
Unless mentioned otherwise, the distances given in the following
sections are calculated by correcting the systemic velocities of the
galaxies for Virgocentric infall (following the formalism presented in
\citealt{mould-2000}), and assuming a Hubble flow with a Hubble
constant of $H_0=75\mathrm{\kms\, Mpc^{-1}}$. Given the different
distance outcomes of different flow models, we use half the difference
between the minimum and maximum flow-corrected distances reported in
NED as an indication of the uncertainty in the distance. Where
independent distance measurements through measurements of the
luminosity of the tip of the red giant branch are available, we adopt
the distances and uncertainties listed in the source paper.

\subsection{Sub-samples}

As will be discussed extensively in Sect.~\ref{sec:thebtf}, we will split our sample
into two sub-samples, namely the ``profile width'' sub-sample and the
``rotation curve'' sub-sample. The profile width sub-sample contains
the galaxies for which it proved impossible to derive a maximum rotation
velocity from either the tilted-ring analysis (\vvf) or from the
position-velocity diagram (\vpv). For this sample, therefore,
only \20\ and \50\ measurements are available.  Their properties are
summarized in Table~\ref{table:btf:derived-params}.  The rotation
curve sub-sample is summarized in Table~\ref{table:btf:rotcur-params}.
It contains the galaxies for which we were able to derive \vvf\ and/or
\vpv\ (in addition to \20\ and \50).

In the Appendix, we show for all galaxies presented in this paper a
summary panel consisting of moment maps, major and minor axis
position-velocity diagrams and the global \HI\ profile. We
additionally show channel maps for the galaxies of the rotation curve
sub-sample.

\begin{figure}[h!]
   \centering
   \includegraphics[width=8cm, bb=97 355 340 565, clip=]{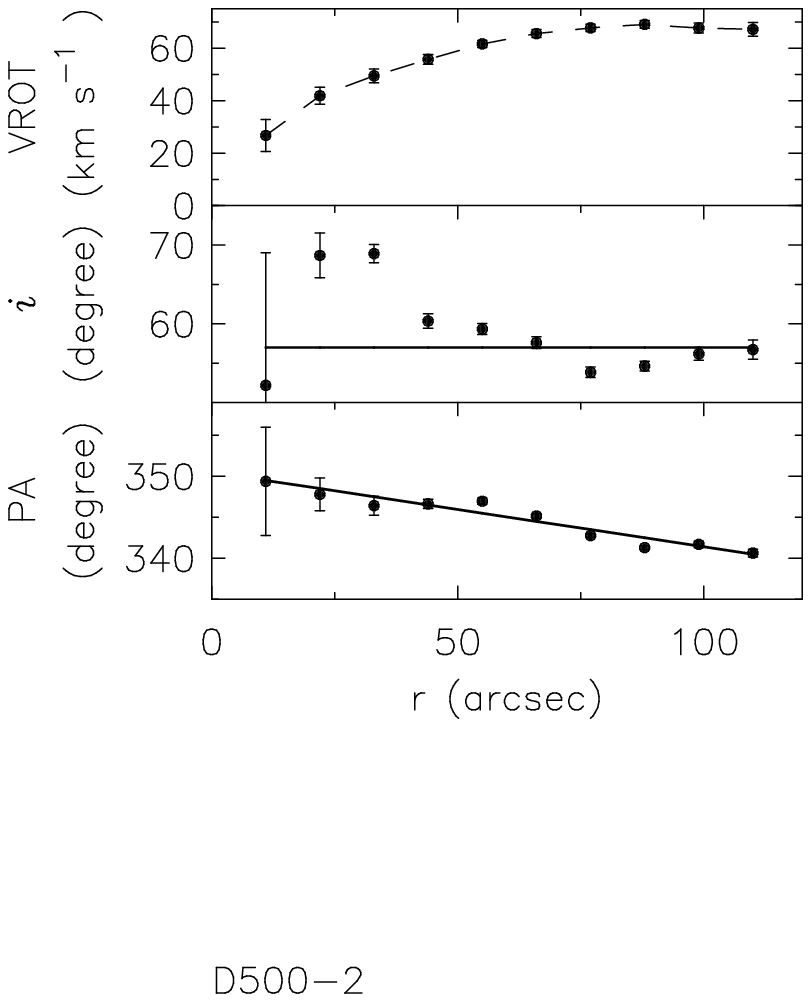}
   \caption[Tilted-ring analysis of D500-2]{Tilted-ring analysis of
     D500-2. From top to bottom, the radial distributions of the
     rotation velocity, the inclination angle, and the position angle
     are shown. The data points in the center and bottom panel
     indicate the values derived for $i$ and PA as free
     parameters. The solid lines indicate the values adopted to derive
     the rotation curve shown in the top panel.  }\label{fig:btf:d500-2-rc}
\end{figure}

\subsection[D500-2]{D500-2 (Data presented in
  Figs. \ref{fig:btf:d500-2-rc}, ~\ref{fig:btf:d500-2}, and
  \ref{fig:btf:d500-2-chan})}
D500-2 is also known as UGC 5716 and is classified as an Sm galaxy.
We assume a distance of $17.9 \pm 3.4$ Mpc.  The global \HI\ profile of D500-2
(cf.~Fig.~\ref{fig:btf:d500-2}) shows the double-horned profile
typical for spiral galaxies, and its velocity field indicates that
D500-2 is clearly dominated by rotation.  We fit ellipses using the
zeroth moment map at a few representative intensity levels.  The
resulting inclination of the \HI\ disk, corrected for the beam size,
is $\ihi \sim57\degr$. The parameters from these isophote fits are
used as initial estimates for the tilted-ring analysis.  The width of
the rings is set to 11\arcsec.  The systemic velocity as derived in
the first tilted-ring fit shows only small radial variation. Its mean
value is $V_{\mathrm{sys}} = 1259\kms$, identical to the the central
velocity of the \50 profile (1259 \kms). Fixing the systemic velocity,
we derive the dynamical center in a new fit by averaging the
(\xpos, \ypos) values over the entire radial range. The resulting
central position (cf. Table~\ref{table:btf:rotcur-params}) is in good
agreement with the optical center from NED and with that derived from our
ellipse fits to the zeroth moment map. Keeping the center position fixed for the
subsequent tilted-ring fits, we derive the PA and inclination angle. The
position angle shows a linear decrease from 350\degr\ in the center to
340\degr\ in the outer parts.

The inclination is then determined in an additional fit by averaging
the inclination values for $r \ge 35\arcsec$.  The resulting kinematic
inclination is $\ikin \sim57\degr \pm 6\degr$, which agrees with the
(beam corrected) inclination of the \HI\ disk. In a last tilted-ring
fit, we derive the rotation velocity by keeping all parameters except
$V_{\rm rot}$ fixed to their previously derived values. After a gentle
inner rise, the rotation curve of D500-2 reaches a flat part at a
velocity of $\vvf \sim 68\kms$. The rotation curve, as well as the
radial distributions for PA and \ikin\ are shown in
Fig.~\ref{fig:btf:d500-2-rc}.  The maximum rotation velocity from the
\emph{pV}-diagram is $\vpv \sim68\kms$.

\subsection[D500-3]{D500-3 (Data presented in Figs.~\ref{fig:btf:d500-3} and \ref{fig:btf:d500-3-chan})}
D500-3 is a dwarf irregular (dI) at a distance of $18.5 \pm 3.2$ Mpc.  Its
velocity field (cf. Fig.~\ref{fig:btf:d500-3}) indicates solid-body
rotation throughout the entire disk, as confirmed by the $pV$-diagram.
We fit isophotes to different intensity levels in the zeroth moment
map and find an inclination of the \HI\ disk $\ihi \sim 42\degr$. This
value is somewhat smaller than the optical \emph{I}-band inclination
of $\iopt \sim 55\degr$ as given in \cite{pildis-1997}.  Because of
the large \HI\ beam size, we consider the optical inclination superior
to the \HI\ inclination and use it in our further analysis.  The
solid-body rotation and the limited spatial resolution make it
impossible to derive a well-determined rotation curve using
tilted-ring models. We are, however, able to get an estimate for the
maximum rotation velocity by using the position-velocity diagram and
find $\vpv \sim 45 \kms$.

\begin{figure}[h!]
   \centering
   \includegraphics[width=8cm, bb=97 355 340 565,
   clip=]{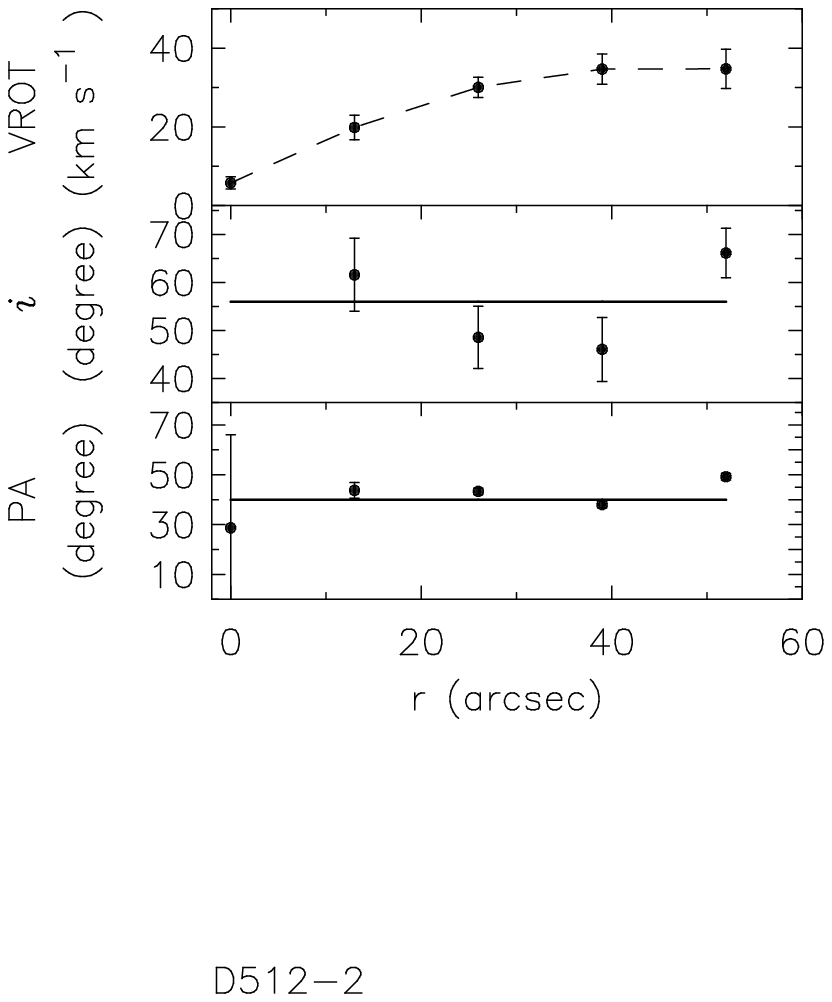}
   \caption[Tilted-ring analysis of D512-2]{Tilted-ring analysis of
     D512-2. The layout is identical to that of
     Fig.~\ref{fig:btf:d500-2-rc}.  }
         \label{fig:btf:d512-2-rc}
   \end{figure}

\subsection[D512-2]{D512-2 (Data presented in
  Figs.~\ref{fig:btf:d512-2-rc}, \ref{fig:btf:d512-2}, and
  \ref{fig:btf:d512-2-chan})}
D512-2 is a galaxy of Hubble type Sm. We assume a distance of $14.1 \pm 2.2$
Mpc.  The velocity field of D512-2 shows signs of solid-body rotation
in its inner parts. We fit ellipses at a number of representative
intensity levels in the zeroth moment map and derive an inclination of
the \HI\ disk of $\ihi \sim 48\degr$ (corrected for the beam
size). Because the apparent size of D512-2 is small, we do not fit the
center position or the systemic velocity using tilted-ring fits. For
the center position, we adopt the estimate derived from the ellipse
fitting, which is in good agreement with the position of the optical
center. For the systemic velocity, we use the central velocity of the
\50\ profile. Keeping the center and the systemic velocity fixed, we
derive an average position angle PA $\sim 40\degr$ from the tilted
ring fit, choosing the width of the tilted-rings to be 13\arcsec.  The
position angle is then kept fixed for a subsequent tilted-ring fit to
estimate $i$. Averaging the resulting inclination values yields
$\ikin\sim 56\degr\pm 10\degr$, which is consistent with the values
discussed above. Using the kinematic inclination, we determine the
rotation curve in a last tilted-ring run leaving only $V_{\rm rot}$
unconstrained.  The rotation curve (cf. Fig.~\ref{fig:btf:d512-2-rc})
confirms what was already suggested by the velocity field: a
solid-body rotation in the inner parts and a flat part in the outer
regions. The rotation velocity of the flat part of the rotation curve
is $\vvf \sim 35\kms$.  From the \emph{pV}-diagram, we estimate the
(inclination corrected) maximum rotation velocity to be $\vpv
\sim 37\kms$.

\begin{figure}[h!]
   \centering
   \includegraphics[width=8cm, bb=97 355 340 565,
   clip=]{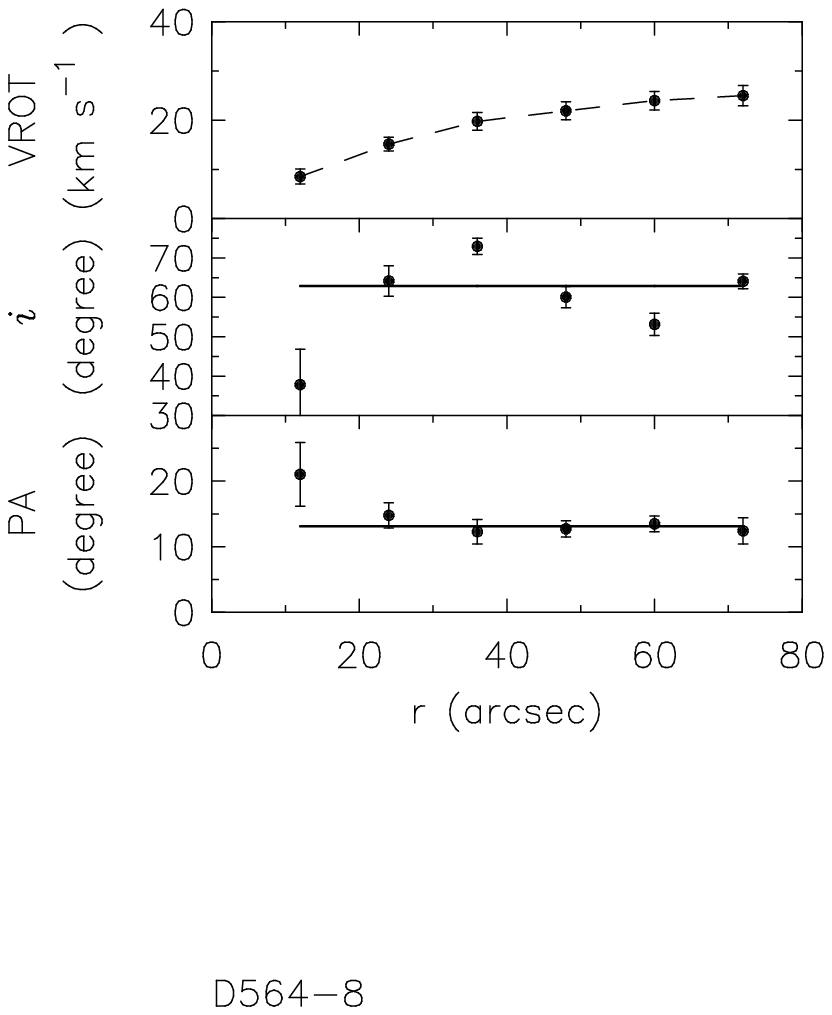}
   \caption[Tilted-ring analysis of D564-8]{Tilted-ring analysis of
     D564-8. The layout is identical to that of
     Fig.~\ref{fig:btf:d500-2-rc}.  }
   \label{fig:btf:d564-8-rc}
   \end{figure}

\subsection[D564-8]{D564-8 (Data presented in
 Figs.~\ref{fig:btf:d564-8-rc}, \ref{fig:btf:d564-8}, and
 \ref{fig:btf:d564-8-chan})}
For the dwarf irregular D564-8 (also known as F564-V3, see
\citealt{deblok-1996b}), we assume a distance of $6.5 \pm 2.6$ Mpc.
The global \HI\ profile of D564-8 is asymmetric with its peak flux
density towards the receding side of the galaxy. The asymmetry of the
global \HI\ profile can be traced also in the \emph{pV}-diagram.  The
maximum rotation velocity, obtained from the \emph{pV}-diagram is
$\vpv \sim29\kms$.  We use ellipse fitting to the zeroth moment map to
derive initial estimates for our tilted-ring fits.  We set the width
of the tilted-rings to 12\arcsec.  In the first tilted-ring fit, we
derive a systemic velocity of $V_{\mathrm{sys}}\approx478 \kms$, which
agrees well with the central velocity of the \50 profile. In a
subsequent fit, we derive the dynamical center by averaging the values
for \xpos\ and \ypos\ over the entire radial range. The resulting
center (cf. Table~\ref{table:btf:rotcur-params}) agrees to within a
few arcseconds with the optical center from NED. In two subsequent
tilted-ring fits, we derive the PA and inclination by averaging the
individual measurements over the radial range, excluding the innermost
(deviant) data point. The resulting inclination ($\ikin \sim63\degr
\pm 7\degr$) is larger than both the one derived from the \HI\ disk
($\ihi \sim50\degr$), and the optical inclination ($\iopt\sim35\degr$,
\citealt{pildis-1997}). However, an inclination of 50\degr, or even
35\degr\ can be ruled out by our kinematic data. Keeping all
parameters except the rotation velocity fixed to their best estimates,
we derive the rotation curve of D564-8. The maximum rotation velocity
is $\vvf\sim 25\kms$ (see Fig.~\ref{fig:btf:d564-8-rc}).

\subsection{D565-5}
The WSRT observations of D565-5 do not detect any emission at the
reported position and redshift.  We do, however, see strong emission
from the nearby galaxy NGC 2903 at the edge of the primary beam. The
agreement of the apparent velocity of this emission with that of the
\cite{schombert-1997} detection suggests that the latter detection was
simply NGC 2903 emission that was picked up with the larger
single-dish beam.

\subsection[D572-5]{D572-5 (Data presented in Fig.~\ref{fig:btf:d572-5})}
D572-5 is an irregular galaxy at a distance of $14.6 \pm 3.2$ Mpc.
Our \HI\ data of this galaxy are barely resolved (see, e.g., the
moment maps in Fig.~\ref{fig:btf:d572-5}), which makes it difficult to
derive a reliable \HI\ inclination.  We therefore adopt the value of
the $I$-band inclination $\iopt\sim 50\degr$ as given by
\cite{pildis-1997} in further analysis.  Due to the low resolution,
substantial off-axis \HI\ emission enters the beam, causing the large
velocity range in the minor axis \emph{pV}-diagram. Given these
problems, a rotation curve or a \vpv\ from the major axis
position-velocity diagram cannot be reliably derived. The only usable
indicators for the maximum rotation velocity are \20\ and \50.

\subsection[D575-1]{D575-1 (Data presented in Fig.~\ref{fig:btf:d575-1})}
D575-1, also known as IC 3810, is classified as an Sm/Irr galaxy. We
assume a distance of $10.0 \pm 3.6$ Mpc. Its global \HI\ profile is
well described by a Gaussian. The first-moment map and the
position-velocity diagrams show only a small velocity range and it
proved impossible to estimate either \vvf\ or \vpv.  The inclination
of the \HI\ disk (derived using ellipse fitting and corrected for the
beam size) is $\ihi \sim53\degr$ and in reasonable agreement with the
optical inclination ($\iopt \sim 61\degr$) from \cite{pildis-1997}
(though the observed small velocity range in the velocity field and
$pV$ diagram casts some doubt on the validity of these values).

\begin{figure}[h!]
   \centering
   \includegraphics[width=8cm, bb=97 355 340 565,
   clip=]{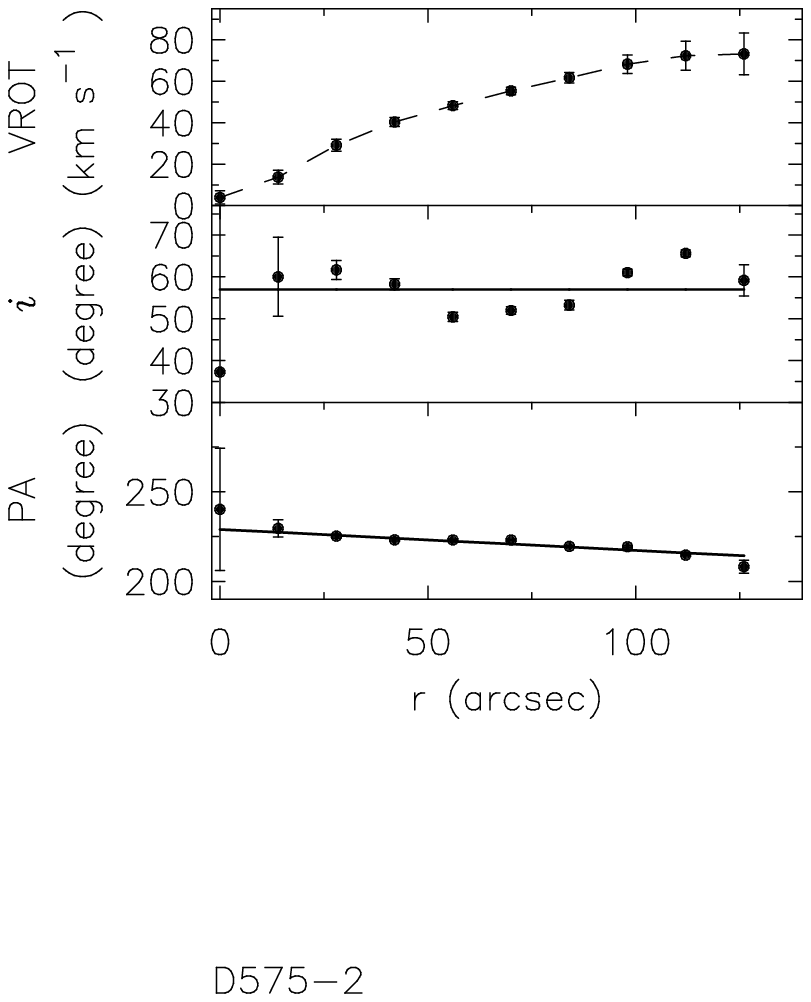}
   \caption[Tilted-ring analysis of D575-2]{Tilted-ring analysis of
     D575-2. The layout is identical to that of
     Fig.~\ref{fig:btf:d500-2-rc}.  }
         \label{fig:btf:d575-2-rc}
   \end{figure}

\subsection[D575-2]{D575-2 (Data presented in
  Figs.~\ref{fig:btf:d575-2-rc}, \ref{fig:btf:d575-2}, and
  \ref{fig:btf:d575-2-chan})}
D575-2, or UGC 8011, is a galaxy of the Hubble type Im, with an
assumed distance of $12.2 \pm 4.5$ Mpc.  It was not possible to estimate \vpv\
from the position-velocity diagram.  The galaxy is kinematically
lopsided, which can be seen in the differences in the velocity
contours between the approaching and receding side. Using ellipse
fitting, we determine the inclination of the \HI\ disk to be $\ihi
\sim 52\degr$, which is somewhat less inclined than what
\cite{pildis-1997} found optically ($\iopt \sim 63\degr$). With the
\HI\ ellipse fitting results as initial estimates for a tilted-ring
analysis and adopting a width of the tilted-rings of 14\arcsec, we
determine a systemic velocity of ($V_{\mathrm{sys}}\sim
774\kms$). This agrees to within 2\kms\ with the center of the \50
profile. Fixing the systemic velocity, we determine a position of the
dynamical center (see Table~\ref{table:btf:rotcur-params}) in
excellent agreement with the optical center as given in NED. Leaving
PA and $i$ unconstrained, we find a gradual decrease of the position
angle from $\sim$\,230$\degr$ in the inner parts to $\sim$\,210$\degr$
in the outskirts of D575-2. Fixing the PA to these values, we derive
the inclination by averaging over all data points with
$r>0\arcsec$. The resulting inclination (\ikin$\sim 57\degr \pm
5\degr$) is halfway between the optical inclination and the one from
the \HI\ disk. Keeping the inclination fixed to the kinematic
estimate, we derive the rotation curve of D575-2 (see
Fig.~\ref{fig:btf:d575-2-rc}). After a linear increase in the inner
parts of the galaxy, the rotation velocity reaches a flat part at
$\sim 74$ \kms.

\subsection[D575-5]{D575-5 (Data presented in Fig.~\ref{fig:btf:d575-5})}\label{sec:D575-5}
D575-5 is classified as a dwarf irregular at an assumed distance of
$7.7 \pm 3.1$ Mpc. Its global \HI\ profile has a Gaussian shape and
its \50 is the smallest of our sample (\50\ $\sim 19\kms$ for the
Hanning smoothed data cube).  \cite{pildis-1997} estimate the optical
inclination to be $\iopt \sim 66\degr$, which is higher than the (beam
corrected) inclination of the \HI\ disk, for which we derive $\ihi
\sim 50\degr$.  However, neither the velocity field, nor the major
axis \emph{pV}-diagram show clear signs of rotation.  We were unable
to derive a maximum rotation velocity from either the velocity field,
or from the major-axis position-velocity diagram.  The second-moment
map shows values of 6--8 \kms\ throughout the entire disk. This
suggests D575-5 is close to face-on and that a relatively large
fraction of the profile width is caused by turbulence. For the
inclination correction of \20\ and \50, we use the inclination of the
\HI\ disk, as the appearance of D575-5 rules out the optical
inclination of $\iopt \sim 66\degr$, though we note that the observed
small velocity range indicates that even this \HI\ value is likely to be an
overestimate of the true inclination.

\begin{figure}[h!]
   \centering
   \includegraphics[width=8cm, bb=97 355 340 565, clip=]{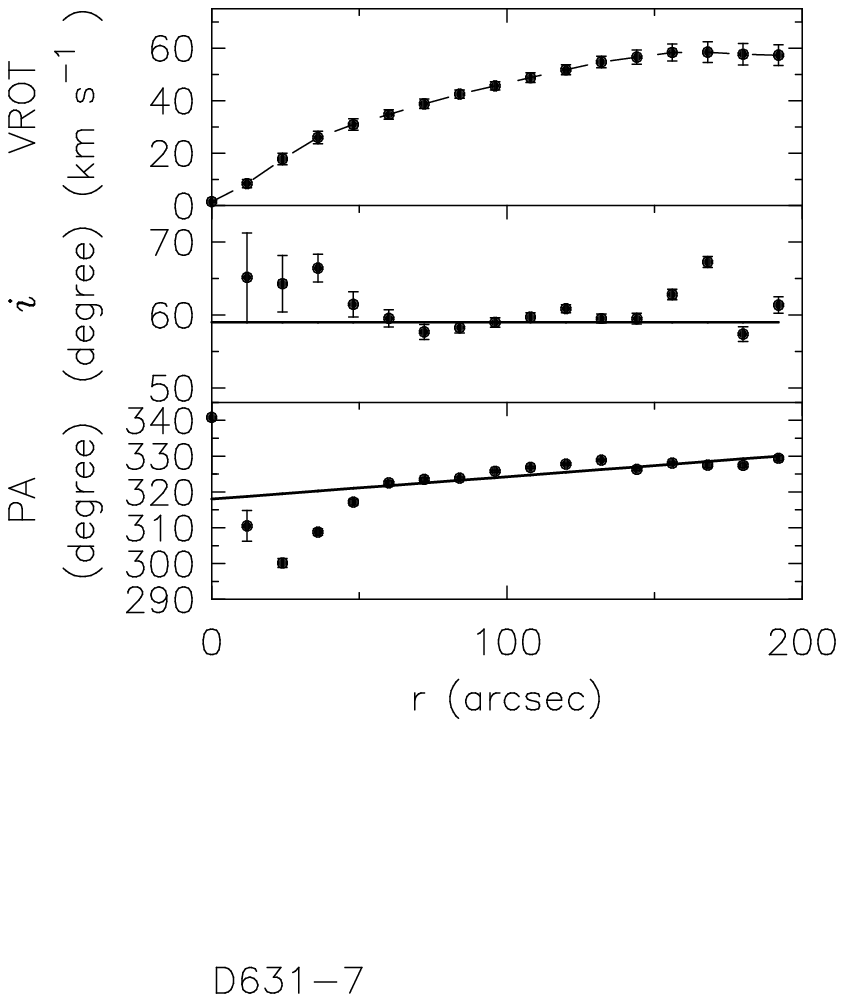}
   \caption[Tilted-ring analysis of D631-7]{Tilted-ring analysis of
     D631-7. The layout is identical to that of
     Fig.~\ref{fig:btf:d500-2-rc}.  }
         \label{fig:btf:d631-7-rc}
   \end{figure}

\subsection[D631-7]{D631-7 (Data presented in Figs.~\ref{fig:btf:d631-7-rc}, \ref{fig:btf:d631-7},  and \ref{fig:btf:d631-7-chan})}
D631-7 is also known as UGC 4115 and is classified as a dwarf
irregular. \cite{karachentsev-2003} estimate its distance to be $5.5 \pm 0.6$
Mpc using the luminosity of the tip of the red giant branch.  The
global \HI\ profile of D631-7 is single-peaked. Its velocity field is
well-resolved and shows clear signs of rotation.  The major axis
\emph{pV}-diagram shows indications of a flat rotation curve in the
outer parts of the galaxy.  Ellipse fitting yields an inclination of
the \HI\ disk $\ihi \sim 57\degr$. The ellipse fit results are used as initial
estimates for the tilted ring analysis.  The width of the rings is set
to 12\arcsec.  In the first tilted-ring fit, we derive a systemic
velocity of $V_{\mathrm{sys}}\sim311\kms$, which agrees well with the
center of the \50 profile. The dynamical center is fitted in the
second run with \rotcur. The resulting center position is given in
Table~\ref{table:btf:rotcur-params}, and agrees to within 1\arcsec\
with the optical center as given by NED. The PA of D631-7 is obtained
in a subsequent fit with \rotcur\ and shows a gradual increase from
$\sim$\,318$\degr$ in the inner parts to $\sim$\,330$\degr$ in the
outer parts. Fixing the PA to these values, we derive the kinematic
inclination of D631-7 to be $\ikin\sim 59\degr \pm 3\degr$ by
averaging the individual tilted-ring values for $50\arcsec \le r \le
150\arcsec$. This is in good agreement with the inclination of the
\HI\ disk ($\ihi \sim 57\degr$), but more face-on than the optical
inclination \citep[$\iopt \sim 66\degr$, derived
by][]{pildis-1997}. Keeping all parameters except \VROT\ fixed, we
determine the rotation curve of D631-7 (see
Fig.~\ref{fig:btf:d631-7-rc}). It shows the typical solid-body
rotation in the inner parts, but reaches a flat part at $\vvf \sim
58\kms$. Assuming the kinematical inclination we derive $\vpv \sim
53\kms$ from the $pV$ diagram.

\subsection[D640-13]{D640-13 (Data presented in Fig.~\ref{fig:btf:d640-13})}
D640-13 is an Im/Sm type galaxy with an adopted distance of $13.4 \pm 3.4$ Mpc.
Its global \HI\ profile has a Gaussian shape. Comparing the size and
the shape of the beam with that of the moment maps shows that the
galaxy is barely resolved.  This heavily affects the derived
inclination of the \HI\ disk ($\ihi \sim 71\degr$) which is much
higher than the optical inclination ($\iopt \sim 48\degr$) by
\cite{pildis-1997}. We therefore use the optical inclination for our
further analysis.  As neither the position-velocity diagrams, nor the
first-moment map show clear signs of rotation, only the widths of the
\20 and \50 profiles remain as a proxy for the maximum rotation
velocity, which we correct using the optical inclination.

\subsection[D646-7]{D646-7 (Data presented in Fig.~\ref{fig:btf:d646-7})}
D646-7 (or UGC 8091) is classified as an irregular dwarf galaxy for
which \cite{karachentsev-2003} estimate a distance of $2.1  \pm 0.6$ Mpc, using
the tip of the red giant branch. The global \HI\ profile of D646-7 has
a clear Gaussian shape. The galaxy lacks clear signs of rotation, both
in the velocity field and in the major-axis position-velocity
diagram. Thus, only \20 and \50 remain as indicators for the maximum
rotation velocity. The optical inclination of D646-7, derived by
\cite{pildis-1997}, is $\iopt \sim 55\degr$, which is in good
agreement with the (uncorrected) inclination of the \HI\ disk ($\ihi
\sim 51\degr$). However, given that the galaxy is not well resolved,
we use the optical inclination in our analysis.

\section{Analysis}\label{btf:sec:analysis}
In the previous section, we have derived several estimates for the
maximum rotation velocities of the galaxies in our sample. Before
using them to construct a BTF, we discuss our choices for the stellar
mass-to-light ratios (\ML), describe the corrections applied to the
line width measurements and estimate the uncertainties in the rotation velocities and
baryonic masses.

\subsection{The choice of  the stellar mass-to-light ratio \ML}\label{sec:btf:choice-ML}
\begin{figure}[]
   \centering
   \includegraphics[width=8cm]{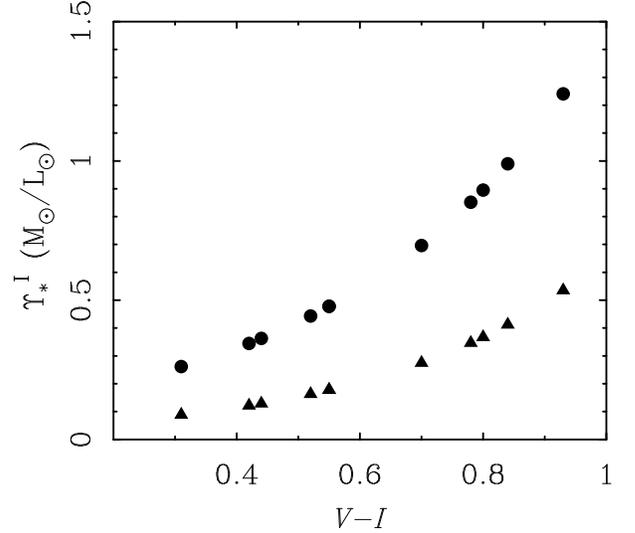}      

   \caption{Stellar {\it I}-band mass-to-light ratio (\ML) {\it vs.}
     $(V-I)$ color for two different population synthesis models. The
     circles are based on the models from \cite{bell-2001} with a
     scaled Salpeter IMF, the triangles are based on the
     \cite{portinari-2004} models with a Kroupa IMF. Note that D631-7
     and D640-13 have the same color: for clarity reasons we only plot
     one point.} \label{fig:btf:comparison-ML-popsynth}
\end{figure}

One of the largest contributors to the scatter in baryonic mass is the
assumed value for the stellar mass-to-light ratio \ML\ and its
uncertainty.  Fortunately, the effect of \ML\ becomes less important
when dealing with low-mass galaxies, as their stellar mass generally
contributes less to the total baryonic mass than the gas mass does
(for reasonable values of \ML).  The \ML\ values adopted here are
derived on the basis of two different population synthesis models
using two different Initial Mass Functions (IMFs). The first model is
from \cite{bell-2001} and uses a scaled Salpeter IMF
\citep{salpeter-1955}. The second population synthesis model discussed
here uses a Kroupa IMF \citep{kroupa-1998} and is taken from
\cite{portinari-2004}.  Figure~\ref{fig:btf:comparison-ML-popsynth}
shows the derived stellar ({\it I}-band) mass-to-light ratios of our
galaxies for the two population models as a function of color.

\begin{figure}[]
   \centering
   \includegraphics[width=8cm, clip=]{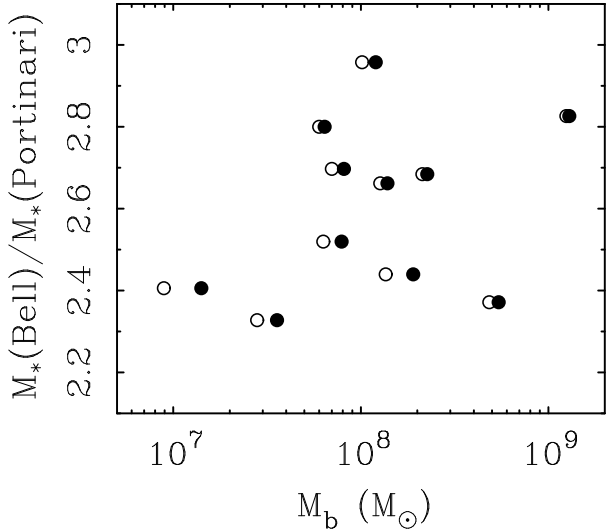}
   \caption{The ratio of the stellar mass from \cite{bell-2001} and
     \cite{portinari-2004} {\it vs.} the implied baryonic mass.
     Baryonic masses derived using the \cite{bell-2001} stellar masses
     are shown as filled circled; baryonic masses derived using the
     \cite{portinari-2004} models as open circles. It is clear the
     choice of \ML\ has only a small effect on the baryonic masses.
   } \label{fig:btf:BTF-diff-ML-ratios}
\end{figure}

It is immediately apparent that the \ML\ from \cite{bell-2001} are
consistently larger than the ones from \cite{portinari-2004}. Although
this difference can amount to a factor close to three, the impact on
the baryonic mass (i.e., the sum of the stellar mass and the gas mass)
is rather small, as can be seen in
Fig.~\ref{fig:btf:BTF-diff-ML-ratios}, where, for each galaxy, we
compare the ratio of the two different stellar mass assumptions
against the baryonic mass implied by the respective \ML\ values.  In
the following analysis we will adopt the average \ML\ values implied
by the two models. The individual mass-to-light ratios are listed in
Table~\ref{table:btf:M/L-ratios}.

\begin{table*}
\begin{minipage}[t]{\linewidth}
\caption[Stellar \ML\ ratios]{Stellar, gas, and baryonic masses.}
\label{table:btf:M/L-ratios}
\centering
\renewcommand{\footnoterule}{}  
\begin{tabular}{lccccccc}
\toprule
\multicolumn{1}{c}{ID}& $\Upsilon_*^I$ (Bell) & $\Upsilon_*^I$ (Portinari) & $\langle \Upsilon_*^I \rangle$ & $M_{\rm{gas}}$ & $M_*$ (Bell) & $M_*$ (Portinari) & $\langle M_{\rm{bar}} \rangle$ \\
\multicolumn{1}{c}{} & & & & $10^8$ \Msun & $10^8$ \Msun & $10^8$ \Msun & $10^8$ \Msun \\

\multicolumn{1}{c}{(1)} & (2) & (3) & (4) & (5) & (6) & (7) & (8) \\
\midrule
D500-2 & 0.35 & 0.12 & 0.24 $\pm$ 0.12 &  12.19	& 0.73	& 0.26	& 12.69\\
D500-3 & 0.26 & 0.09 & 0.18 $\pm$ 0.09 &  0.92	& 0.28	& 0.09	& 1.11\\
D512-2 & 0.90 & 0.37 & 0.64 $\pm$ 0.27 &  0.98	& 0.93	& 0.38	& 1.63\\
D564-8 & 1.24 & 0.54 & 0.89 $\pm$ 0.35 &  0.22	& 0.14	& 0.06	& 0.32\\
D572-5 & 0.44 & 0.16 & 0.30 $\pm$ 0.14 &  1.20	& 0.19	& 0.07	& 1.33\\
D575-1 & 0.70 & 0.28 & 0.49 $\pm$ 0.21 &  0.53	& 0.26	& 0.10	& 0.71\\
D575-2 & 1.07 & 0.45 & 0.76 $\pm$ 0.31 &  4.41	& 1.02	& 0.43	& 5.13\\
D575-5 & 0.36 & 0.13 & 0.25 $\pm$ 0.12 &  0.60	& 0.04	& 0.02	& 0.63\\
D631-7 & 0.48 & 0.18 & 0.33 $\pm$ 0.15 &  2.05	& 0.20	& 0.08	& 2.19\\
D640-13 & 0.48 & 0.18 & 0.33 $\pm$ 0.15 & 0.64	& 0.18	& 0.07	& 0.76\\
D646-7 & 0.99 & 0.41 & 0.70 $\pm$ 0.29 &  0.05	& 0.09	& 0.04	& 0.12\\
\bottomrule
\end{tabular}
\end{minipage}
\newline
{\sc Notes: }
(1): galaxy identifier; (2): stellar {\it I}-band mass-to-light ratio based 
on \cite{bell-2001}; (3): stellar {\it I}-band mass-to-light ratio based on 
\cite{portinari-2004}; (4): average of {\it I}-band \ML\ values listed in columns (2) and (3).  
For the uncertainty in \ML, we assume half the difference between the mass-to-light ratios of 
the two population models; (5): the gas mass in $10^8$ \Msun; (6): the stellar mass based on \ML\ 
in column (2) in units of $10^8$ \Msun; (7): like (6), but based on \ML\ in column (3); 
(8): the baryonic mass used in further analysis, derived from the gas mass in 
column (5) and the average of the stellar masses in columns (6) and (7).
\end{table*}

\subsection{Line width corrections}\label{sec:btf:lw-corrections}

The rotation velocities derived using the profile widths \20\ and  \50,
still need to be corrected for instrumental velocity resolution and turbulence.
To correct for the instrumental resolution, we use the approach of
\citet[][]{verheijen-1997}:
\begin{equation}
W_{x}=W_{x}^{\mathrm{obs}}-C_{x}\cdot \left [ \sqrt{1+\left ( \frac{R}{23.5}\right ) ^2}-1 \right ],
\label{eq:btf:instr-res}
\end{equation}
where the subscript $x$ refers to the chosen profile width measure (i.e.,
$x=20$ for \20\ and $x=50$ for \50), $R$ is the instrumental
resolution in \kms (cf.\ col.\ (9) in Table
\ref{table:btf:mapping-params}), and $C_x$ is a constant, equal to
$C_{20}=35.8$ for the \20\ profile and $C_{50}=23.5$ for the \50\
profile. The `obs' superscript in $W_{x}^{\mathrm{obs}}$ denotes the
observed profile width (given in cols.\ (9) and (10) of Tables
\ref{table:btf:derived-params} and \ref{table:btf:rotcur-params}).

In addition to broadening due to finite velocity resolution, we correct
the velocity widths for broadening due to turbulent motion of the
neutral gas.  Following \citet{tully-1985}, we use
\begin{equation}
W^2_{x, {\mathrm{turb}}}=W^2_{x}+W^2_{t,x}\left [ 1-2e^{-\left (
    \frac{W_{x}}{W_{c,x}}\right ) ^2} \right ] -2W_{x}W_{t,x}
    \left [ 1-e^{-\left ( \frac{W_{x}}{W_{c,x}}\right )^2} \right ].
\label{eq:btf:turb-mot}
\end{equation}
The subscript $x$ again refers to the profile width measured at 20 or
50 percent of the peak flux; $W_x$ indicates velocity widths already
corrected for instrumental resolution following
Eq.~\ref{eq:btf:instr-res}.  The $W_{c,x}$ values represent the
typical velocity widths where the shapes of the velocity profiles
change from boxy (double-horned) to Gaussian; the constants $W_{t,x}$
indicate the amount by which turbulent motion of the neutral gas
broadens the \HI\ profile.  For the choice of $W_{c,x}$ and $W_{t,x}$,
we follow \cite{verheijen-1997}, who assumes a turbulent motion of the
gas with a velocity dispersion of 10 \kms\ and values of
$W_{t,20}=22\, \kms$, $W_{t,50}=5\,\kms$, $W_{c,20}=120\, \kms$ and
$W_{c,50}=100\,\kms$ (see Chapter 5 of \citealt{verheijen-1997} for an
extensive discussion of these correction terms).  The resulting
corrected profile widths are given in Cols. (11) and (12) of
Tables~\ref{table:btf:derived-params} and
\ref{table:btf:rotcur-params}. Note that the profile widths listed in
Tables \ref{table:btf:derived-params} and
\ref{table:btf:rotcur-params} are not corrected for inclination.

\subsection{Sources of uncertainties}\label{sec:btf:uncertainties}

\subsubsection{Uncertainty in $V_{\mathrm{max}}$}
The uncertainty in $V_{\mathrm{max}}$ depends on the measurement error
in the observed velocity and that in the inclination value.  For the
velocity uncertainty we assume that we can measure the velocity with
an accuracy of one channel width $(\sim 4 \kms)$.  In addition, we
assume that noise in the data or intrinsic variance within the
galaxies introduce another channel width worth of uncertainty. The total
uncertainty we assume is the quadratic sum of these two terms or $6 \kms$.

For the uncertainty in the derived inclination angle, we distinguish
between galaxies with a kinematically derived inclination, and those
where we had to use the optical inclination or that of the \HI\ disk.
For the former category, we adopt the scatter of the inclination
values of the individual tilted-ring models (see middle panels in
Figs.~\ref{fig:btf:d500-2-rc}-\ref{fig:btf:d631-7-rc}) as the
inclination uncertainty (see Col. (13) of Table
\ref{table:btf:rotcur-params}).  The average value of the
uncertainties in these kinematically derived inclinations is $6\degr$
and is used as the uncertainty in the inclination angles of the
remaining galaxies.  The total uncertainty of the
(inclination-corrected) maximum rotation velocity is then calculated
using Gaussian error propagation.

\begin{figure*}[]
   \centering
   \includegraphics[width=\textwidth]{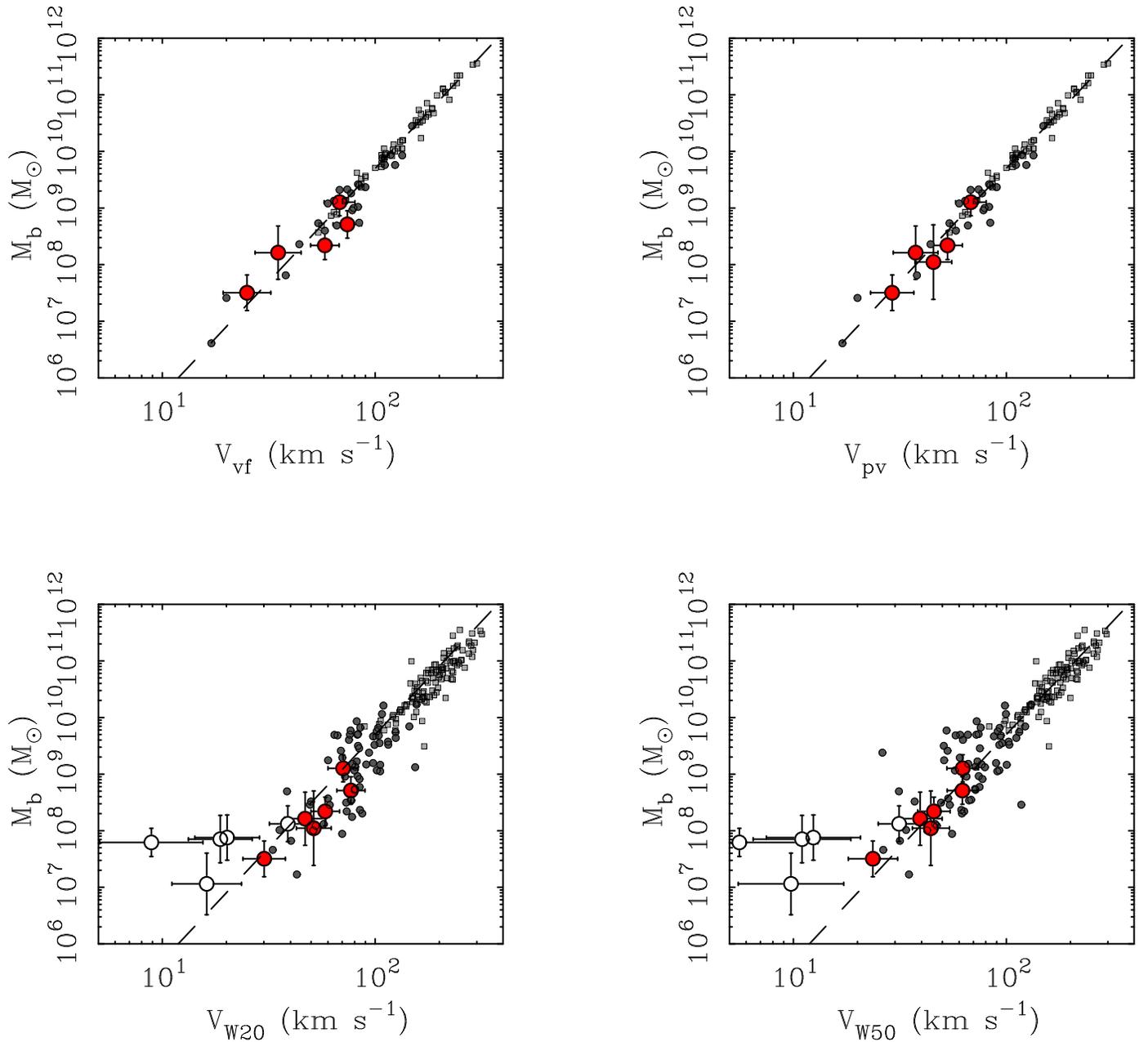}
   \caption[The baryonic Tully-Fisher relation using different
   estimates for $V_{\mathrm{max}}$]{The baryonic Tully-Fisher
     relation using four different estimates for $V_{\mathrm{max}}$,
     namely \vvf\ from the tilted-ring analysis of the velocity fields
     (top left), \vpv\ from the major-axis position-velocity diagram
     (top right), $V_{W20}$, i.e., half the \20\ (bottom left), and
     $V_{W50}$, i.e., half the \50\ (bottom right).  As a reference,
     we show the BTF relation of the analysis of \cite[][dashed
     line]{mcgaugh-2005}. In the top panels, the star-dominated
     galaxies of \citet{mcgaugh-2005} are shown as small squares and
     the gas-dominated galaxies of \citet{stark-2009} are shown as
     small circles.  In the lower panels, the velocity widths from
     \citet{bothun-1985} (small squares) and \citet{mcgaugh-2000} (small circles)
     are shown for comparison.  Our new data are show as large
     circles.  Galaxies for which no rotation is discernable are shown
     as open circles (with line-widths only) and those with rotation
     detected in the position-velocity diagram are shown as filled
     circles.  The latter fall along the extrapolation of the BTF
     relation found by \citet{mcgaugh-2005} while some of the former
     fall off the relation.  These galaxies may not be supported
     by rotation, may have incorrect inclinations, or could simply reflect the increased scatter
     in the relation when line-width is used as a proxy for rotation
     velocity.  }
         \label{fig:btf:BTF}
   \end{figure*}

\subsubsection{Uncertainty of the baryonic mass}
The baryonic mass is the sum of the stellar mass and the gas mass:
\begin{equation}
 M_{\mathrm{bar}}=M_{\mathrm{stars}}+M_{\mathrm{gas}}.
\label{eq:btf:Mbar}
\end{equation}
The stellar mass is calculated through
\begin{equation}
 M_{\mathrm{stars}}=\Upsilon^I_* \, 10^{-0.4[M_V-(V-I)-4.02]}, 
\label{eq:btf:Mstars}
\end{equation}
where $\Upsilon^I_*$ is the stellar mass-to-light ratio in the {\it
  I}-band, $M_I= M_V-(V-I)$ is the absolute {\it I}-band magnitude,
and 4.02 is the magnitude of the Sun in the {\it I}-band.  The
absolute magnitude of a galaxy depends through the distance modulus on
its apparent magnitude and its distance.  Thus, the stellar mass
depends on three quantities: stellar mass-to-light ratio, distance,
and apparent magnitude. For the uncertainty in the mass-to-light
ratio, we use the values derived in Section~\ref{sec:btf:choice-ML}.
For the uncertainty in the distances, we use the values listed in
Sect.~\ref{btf:sec:indiv-galaxies}.  The apparent magnitudes of the
galaxies in our sample were determined by \cite{pildis-1997}, who
report a photometric accuracy of 0.05 mag. This is insignificant
compared to the influence of the uncertainties in the distance and in
the stellar \ML, and we therefore ignore the uncertainties of $m_V$
for the uncertainty of the stellar mass.

The second term contributing to the baryonic disk mass is
$M_{\mathrm{gas}}$, the mass of the gas, which is given by:
\begin{equation}
  M_{\mathrm{gas}}=1.4 M_{HI} = 1.4 \cdot 2.36 \cdot 10^5 D^2 \int S dv, 
\label{eq:btf:Mgas}
\end{equation}
where $M_{HI}$ is the total \HI\ mass, $D$ is the distance in Mpc, $S$
is the total flux in $\mathrm{mJy\,beam^{-1}}$, and $dv$ is the
velocity resolution in \kms.  The constant factor 1.4 corrects the
\HI\ mass for the presence of helium and metals. Note that since
H$_2$/\HI, the ratio between the molecular and the neutral hydrogen,
is much lower in dwarf galaxies compared to luminous spirals
\citep[][]{taylor-1998, leroy-2005}, we do not apply correction terms
to account for molecular hydrogen.  The uncertainty of the \HI\ mass
depends on the uncertainty of the total flux, and quadratically on the
distance uncertainty. We focus again on the uncertainty in the
distance, which is the dominant source of uncertainty here.
Inserting Eqs.~\ref{eq:btf:Mstars} and \ref{eq:btf:Mgas} into
Eq.~\ref{eq:btf:Mbar}, we determine the uncertainty of the baryonic
mass by assuming a Gaussian error propagation of the individual
uncertainties discussed above. The stellar, gas, and baryonic masses
are also listed in Table~\ref{table:btf:M/L-ratios}.

\subsection{The baryonic Tully-Fisher relation}\label{sec:thebtf}
In this section we present the baryonic Tully-Fisher relation for the
galaxies of our sample, using the different estimates for
$V_{\mathrm{max}}$ derived previously.  As a reference, we use the
work by \cite{mcgaugh-2005}.  It presents the BTF
for galaxies with well-determined rotation velocities between
50\,\kms\ and 300\,\kms, and tests several methods to determine
stellar mass-to-light ratios. The one yielding the smallest scatter in
the BTF is based on the mass-discrepancy-acceleration relation
\citep[MDAcc, see][]{mcgaugh-2004} and gives a BTF relation 
\begin{equation}
  M_{\mathrm{bar}}=50 V_{\mathrm{max}}^4,
\label{eq:btf:BTF-mcgaugh}
\end{equation}
with the baryonic mass in $M_{\sun}$ and the maximum rotation velocity in \kms.

For the construction of a BTF relation using our galaxies, we use the
four different estimates for $V_{\mathrm{max}}$ derived earlier. To
recapitulate, these are:
\begin{itemize}
 \item[] (a) \vvf, obtained from a tilted-ring analysis of the velocity fields,
 \item[] (b) \vpv, obtained from fitting the outer parts in major-axis
   position-velocity diagrams,
 \item[] (c) $V_{W20}$, obtained from $\frac{1}{2}\20$, the rotation velocity as
   derived from half the (corrected) width of the \HI\ profile at the
   20 percent level of the maximum flux,
 \item[] (d) $V_{W50}$, obtained from $\frac{1}{2}\50$, the rotation velocity as
   derived from half the (corrected) width of the \HI\ profile at the
   50 percent level of the maximum flux,
\end{itemize}
These velocities are all corrected for inclination. Additional
corrections have been applied to $V_{W20}$ and $V_{W50}$ (as described
in Section~\ref{sec:btf:lw-corrections}).

In Fig~\ref{fig:btf:BTF}, we show the baryonic Tully-Fisher relation
for the galaxies of our sample using the different velocity
estimates. The uncertainties are discussed in
Section~\ref{sec:btf:uncertainties}.  In Fig~\ref{fig:btf:BTF}, we
distinguish between galaxies in which clear signatures of rotation
were detected in the position-velocity diagram and/or the velocity
field (\vpv\ and/or \vvf\ available in addition to $V_{W20}$ and
$V_{W50}$; the \emph{rotation curve sub-sample}), and galaxies for which no
clear rotation was detectable (only $V_{W20}$ and $V_{W50}$ available;
the \emph{profile-width sub-sample}). For all four measures of $V_{\rm max}$,
the galaxies in the rotation curve sub-sample are consistent with the
BTF as derived in \citet{mcgaugh-2005} and \citet{stark-2009}.  

A few galaxies in the profile-width sub-sample are also consistent
with a line-width-based BTF, but the majority of these galaxies are
found to the left of the extrapolated BTF. This may indicate an
increased scatter at the low line-width end of a line-width based
BTF. Alternatively, the galaxies may be more face-on than indicated by
the optical or \HI\ inclinations (e.g., D575-5), or they may not be
supported by rotation. Without independent and/or resolved measures of
the rotation velocity and the kinematic inclinations, it is
difficult to say anything further on these galaxies.

In the following we therefore restrict our analysis to those galaxies
which have a well resolved velocity field which allowed us to derive a
rotation curve or --- at the very least --- a maximum rotation
velocity using the position-velocity diagram (consistent with the
procedure used in \citealt{mcgaugh-2005} and \citealt{stark-2009}). We
emphasize that this is our \emph{only} selection criterion. We do not
preferentially select against galaxies that are not on the BTF, (in
fact, one of the profile-width-only galaxies (D572-5) is right on the
BTF), and we include all galaxies for which \vpv\ or \vvf\ could be
determined regardless of their position in the BTF diagram.

\subsection{The scatter of the BTF}\label{sec:btf:scatter}
The panels in Fig.~\ref{fig:btf:BTF} clearly show that the data points
from our rotation curve sample agree well with the BTF relation from
\cite{mcgaugh-2005} as given in Eq.~\ref{eq:btf:BTF-mcgaugh}.  This
relation was derived using resolved observations of the flat parts of
the rotation curves of the sample galaxies.  Including our five
galaxies with measured values of \vvf\ yields a revised BTF of the
form
\begin{equation}
 M_{\mathrm{bar}}=58 V_{\mathrm{max}}^{3.97}.
\label{eq:btf:BTF-trachternach}
\end{equation}
This is remarkably close to the revised BTF presented in
\citet{stark-2009}, which extends the \citet{mcgaugh-2005} BTF with an independent
sample of low-mass gas-rich galaxies. They find
\begin{equation}
 M_{\mathrm{bar}}=62 V_{\mathrm{max}}^{3.94}.
\label{eq:btf:BTF-stark}
\end{equation}
The three fits given in Eqs.\ \ref{eq:btf:BTF-mcgaugh}
, \ref{eq:btf:BTF-trachternach} and \ref{eq:btf:BTF-stark} are
statistically identical.

For comparison, \citet{bell-2001} find a slope of 3.5 for a
subsample of the \citet{mcgaugh-2005} sample, but choose different
\ML\ values than \cite{mcgaugh-2005} does. Using identical \ML\ values
yields identical slopes. Work by \citet{geha-2006} gives a slope of
3.7, \citet{derijcke-2007} find a slope of 3.15, but this includes
early-type galaxies. An analysis by \citet{meyer-2008} of \HI-selected
galaxies gives a steeper slope of 4.35. The majority of these studies
deal with massive galaxies, dominated by stars, and are therefore
sensitive to assumptions on \ML. The new data presented here provide a
test of the slope extrapolated from the fit to higher mass galaxies
that do not overlap with this sample, and prefer a slope of $\sim
4$. This is also the conclusion of \citet{stark-2009} who calibrate
the BTF using only gas-dominated galaxies, independent of assumptions
on \ML. They find a slope of 3.94 (Eq.\ \ref{eq:btf:BTF-stark}). If a
single, unbroken BTF exists, this can only imply that the slope at the
high mass end must also be close to 4.

We now return to the conclusions of \cite{franx-1992}, who use
the scatter of the TF relation to put constraints on the ellipticity
of dark matter halos. They argue that if the potential in the plane of
the disk is elongated, then the different viewing angles will cause
scatter in the TF relation. In return, the scatter of the (B)TF
relation can be used to put an upper limit on the ellipticity of the
potential in the plane of the disk. According to \cite{franx-1992}, a
TF relation with a scatter of 0.31 mag (0.46 mag when photometric
inclinations are used) constrains the ellipticities of the potentials
to be below 0.10. As it is unlikely that all the scatter in the TF
relation is due to different viewing angles, they argue that an
elongation between 0 and 0.06 is reasonable.

Since the work of \cite{franx-1992}, the quantity and quality of the
data have improved significantly and we can now trace the BTF relation
over a large range of galaxy masses and rotation velocities.  The
scatter in the optimum BTF from \cite{mcgaugh-2005} is $\sigma=0.098$
dex, or 0.25 mag.  Including our galaxies results in a slightly larger
scatter ($\sigma=0.13$ or $\sigma_M=0.33\ \mathrm{mag}$) --- mainly
because the distances to our galaxies are more uncertain.  This
scatter is only marginally larger than the scatter on which
\cite{franx-1992} based their upper limits for the ellipticity of the
potentials in the disk plane.

Their arguments equally apply to our sample. As the small scatter in
the BTF holds down to rotation velocities of $\sim 25 \kms$, it
follows that the elongation of the potentials of these galaxies cannot
be very large, and similar limits as those derived by
\cite{franx-1992}, also apply to our galaxies.  The upper limit on
elongation of the potential of 0.06 is consistent with the results
derived by \cite{trachternach-08} who use harmonic decompositions of
the velocity fields of 18 nearby well-resolved galaxies from the
THINGS survey \citep{walter-07, deblok-07} to put constraints on the
elongation of the potential in the plane of the disk.  They find that
the the average elongation of the potential is small ($0.017\pm
0.020$), particularly when compared to what is found in CDM
simulations \citep[cf.][]{frenk-1988, hayashi-2006}.
\cite{trachternach-08} also find that the elongation does not increase
towards the center of the galaxies.  The tightly correlated BTF
relation presented in this paper constrains the ellipticity of galaxy
potentials using a different method, but reaches similar results.

\section{Conclusions}\label{btf:sec:conclusions}

We present new \HI\ observations of a sample of low-mass dwarf
galaxies and use these to explore and extend the baryonic Tully-Fisher
(BTF) relation at low rotation velocities and galaxy masses.  We
present and discuss several estimates for $V_{\mathrm{max}}$, the
maximum rotation velocity. For galaxies where a clear rotation signal
could be detected, the different estimates are in good agreement and
the BTF relations based on them are equally well-constrained.  We
discuss the choice of stellar mass-to-light ratio (\ML) and show that
its choice is not crucial for the extreme dwarf galaxies in our
sample, since for them, the stellar mass generally contributes less to
the total baryonic mass than is the case for luminous high-mass
galaxies.

The small scatter in the BTF presented here ($\sigma_M=0.33\
\mathrm{mag}$) puts strong constraints on the ellipticity of the
potential in the plane of the disk of the galaxies.  Our results are
in agreement with those from \cite{franx-1992}, \cite{franx-1994}, and
\cite{trachternach-08}, indicating that, at least in the plane of the
disk, galaxy halos are not very elongated.  The small scatter in the
BTF over almost 5 orders of magnitude of baryonic mass indicates it is
a fundamental relation which tightly couples the visible baryonic
matter and the dark matter content of galaxies.

\begin{acknowledgements}
  CT would like to thank Janine van Eymeren and Volker Knierim for
  many stimulating discussions. We also thank the anonymous referee
  for valuable comments.  The work of CT is supported by the German
  Ministry for Education and Science (BMBF) through grant 05 AV5PDA/3.
  The work of WJGdB is based upon research supported by the South
  African Research Chairs Initiative of the Department of Science and
  Technology and National Research Foundation.  The Westerbork
  Synthesis Radio Telescope is operated by the ASTRON (Netherlands
  Foundation for Research in Astronomy) with support from the
  Netherlands Foundation for Scientific Research (NWO).  This research
  has made use of the NASA/IPAC Extragalactic Database (NED) which is
  operated by the Jet Propulsion Laboratory, California Institute of
  Technology, under contract with the National Aeronautics and Space
  Administration.
\end{acknowledgements}
\bibliographystyle{aa}
\bibliography{11136}


\begin{appendix}
\section{Atlas}\label{btf:sec:appendix-atlas}
\begin{figure*}[]
   \centering
   \includegraphics[width=16cm, bb=43 420 560 755, clip=]{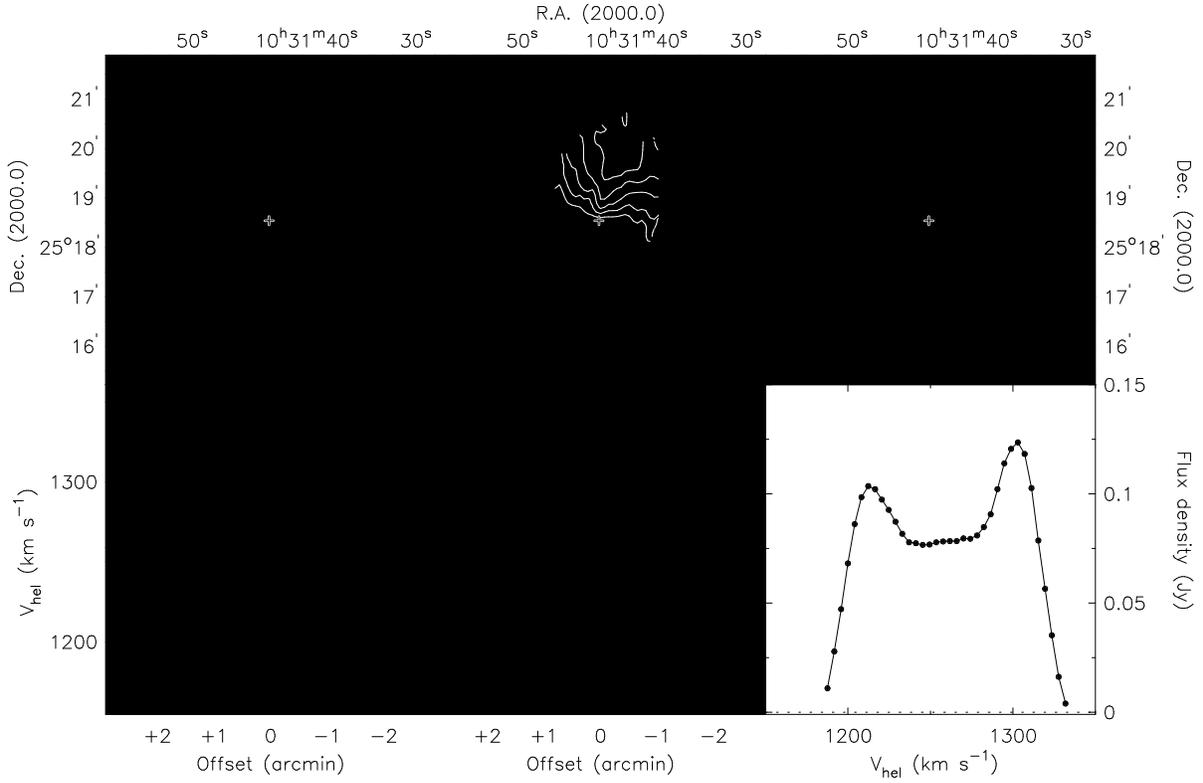}
   \caption[Summary panel for D500-2]{Summary panel for D500-2. A
     general description of the panels and levels is given at the
     beginning of the Appendix.  The $3\sigma$ contour in the zeroth
     moment map corresponds to a flux of 5.09 mJy, which translates to
     a column density limit of $n_{HI}=8.8\cdot10^{19}\
     \mathrm{cm}^{-2}$.  }
         \label{fig:btf:d500-2}
   \end{figure*}

\begin{figure*}
   \centering
   \includegraphics[width=16cm, bb=43 420 560 755, clip=]{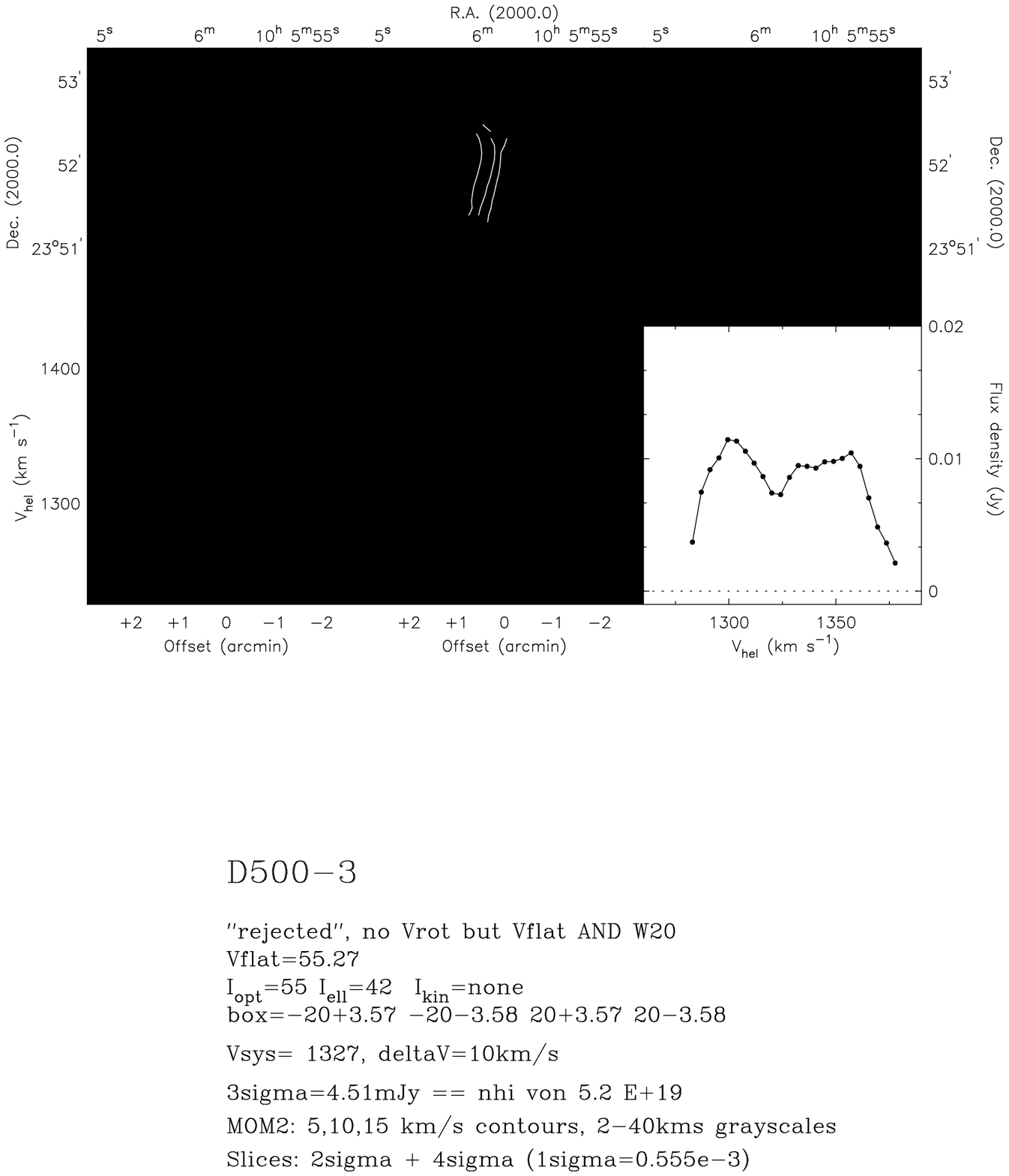}
   \caption[Summary panel for D500-3]{Summary panel for D500-3. A
     general description of the panels and levels is given at the
     beginning of the Appendix. The $3\sigma$ contour in the
     zeroth moment map corresponds to a flux of 4.51 mJy,
     which translates to a column density limit of
     $n_{HI}=5.2\cdot10^{19}\ \mathrm{cm}^{-2}$.  }
         \label{fig:btf:d500-3}
   \end{figure*}

\begin{figure*}
   \centering
   \includegraphics[width=16cm, bb=43 420 560 755, clip=]{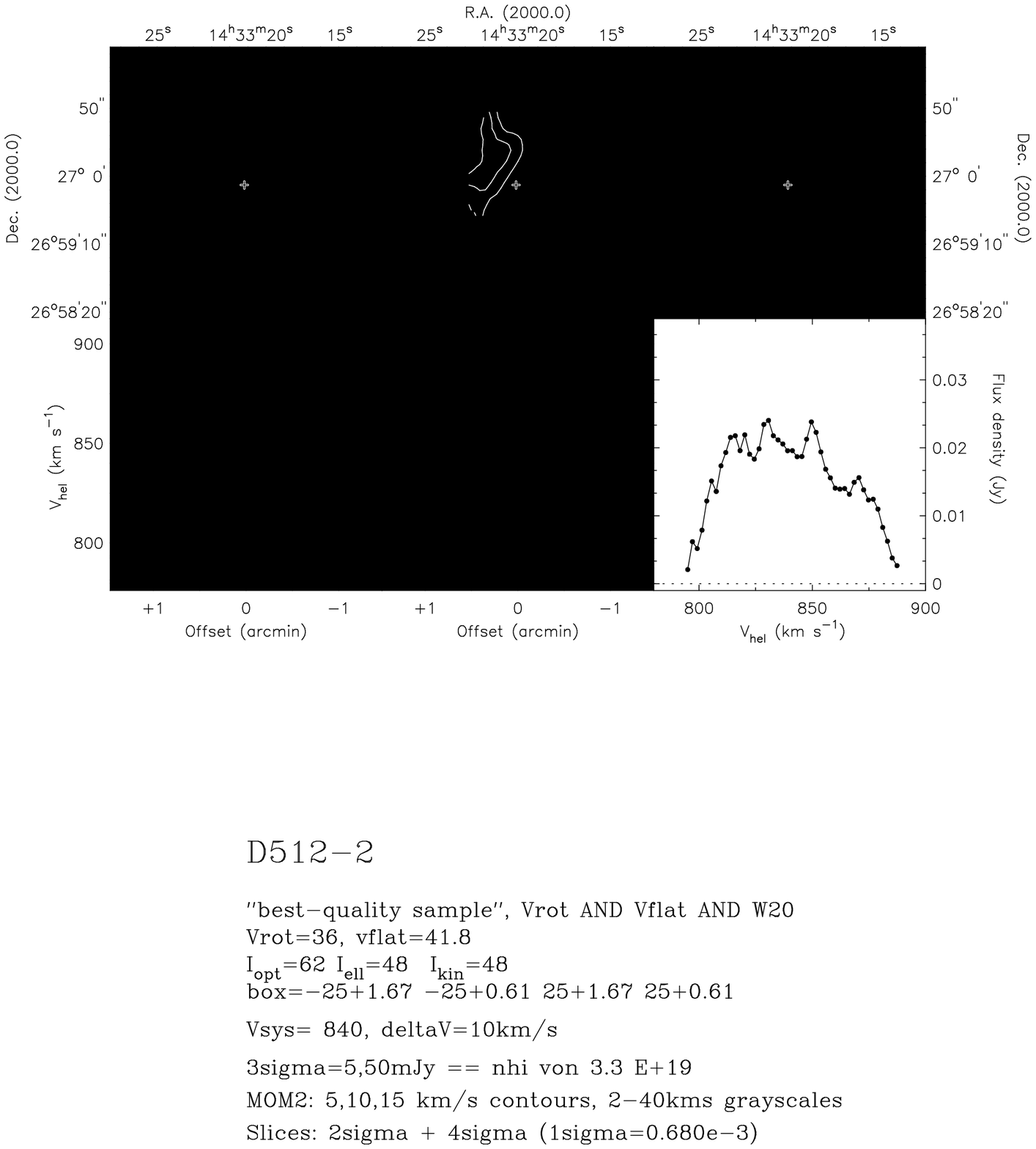}
   \caption[Summary panel for D512-2]{Summary panel for D512-2. A
     general description of the panels and levels is given at the
     beginning of the Appendix. The $3\sigma$ contour in the zeroth
     moment map corresponds to a flux of 5.50 mJy, which translates to
     a column density limit of $n_{HI}=3.3\cdot10^{19}\
     \mathrm{cm}^{-2}$.  }
         \label{fig:btf:d512-2}
   \end{figure*}

\begin{figure*}
   \centering
   \includegraphics[width=16cm, bb=43 420 560 755, clip=]{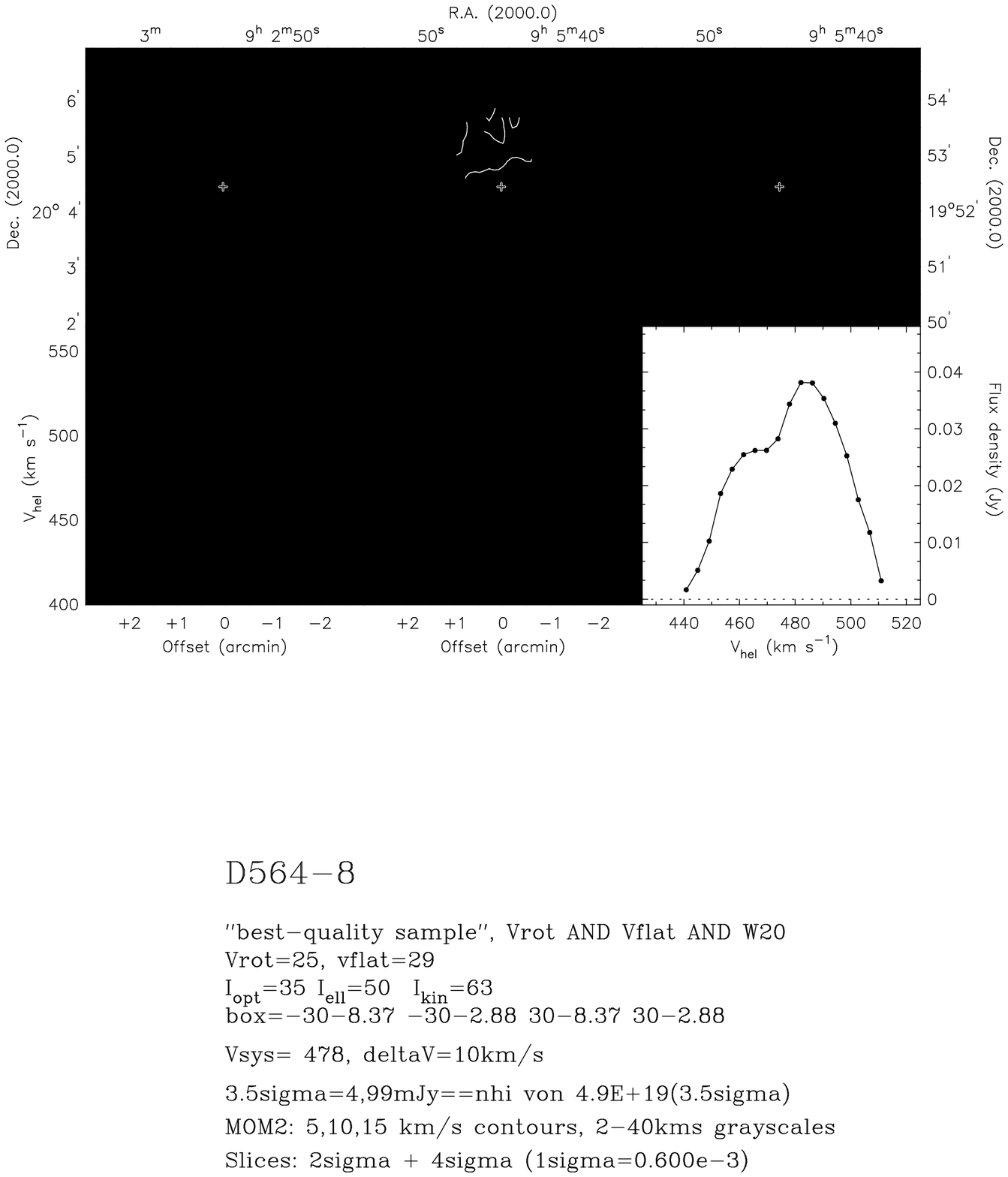}
   \caption[Summary panel for D564-8]{Summary panel for D564-8. A
     general description of the panels and levels is given at the
     beginning of the Appendix. The $3\sigma$ contour in the
     zeroth moment map corresponds to a flux of 4.99 mJy,
     which translates to a column density limit of
     $n_{HI}=4.9\cdot10^{19}\ \mathrm{cm}^{-2}$.  }
         \label{fig:btf:d564-8}
   \end{figure*}

\begin{figure*}
   \centering
   \includegraphics[width=16cm, bb=43 420 560 755, clip=]{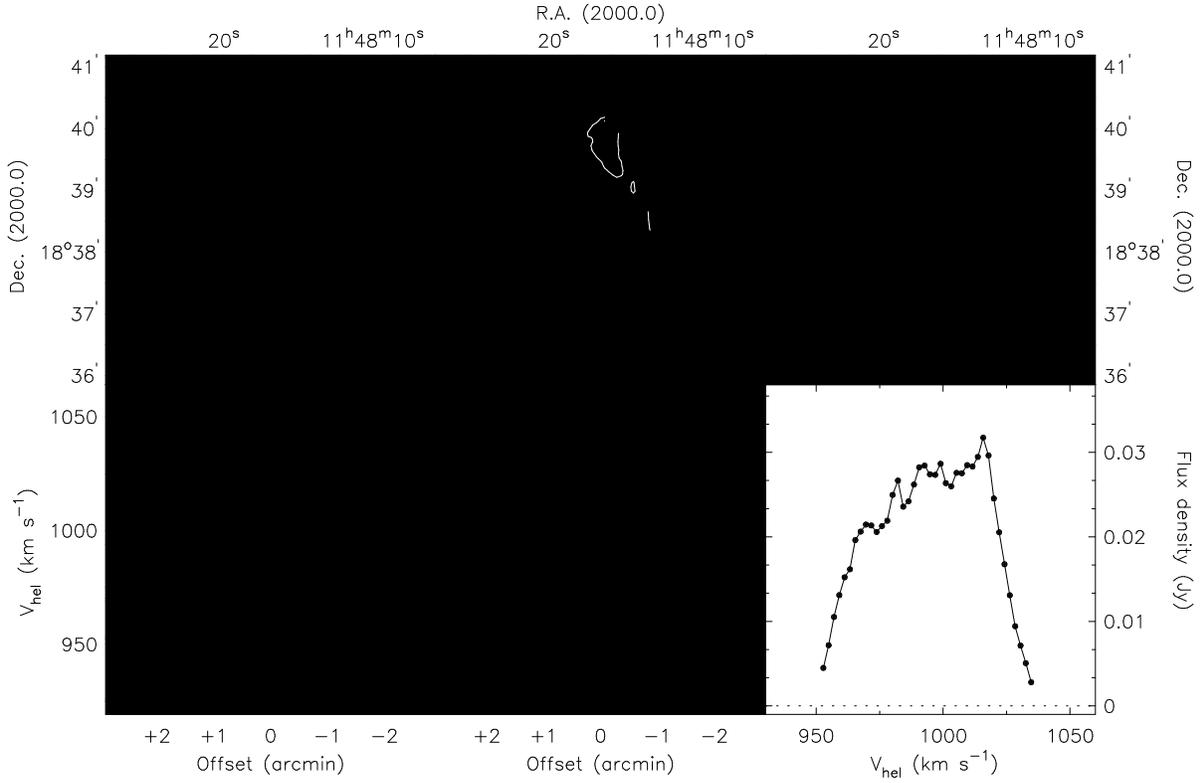}
   \caption[Summary panel for D572-5]{Summary panel for D572-5. A
     general description of the panels and levels is given at the
     beginning of the Appendix. Grayscales in the zeroth moment map
     run from a column density of $n_{HI}=1\cdot10^{19}\
     \mathrm{cm}^{-2}$ (white) to $n_{HI}=8\cdot10^{20}\
     \mathrm{cm}^{-2}$ (black).  The $3\sigma$ contour in the zeroth
     moment map corresponds to a flux of 12.47 mJy, which translates
     to a column density limit of $n_{HI}=1.4\cdot10^{19}\
     \mathrm{cm}^{-2}$.  }
         \label{fig:btf:d572-5}
   \end{figure*}

\begin{figure*}
   \centering
   \includegraphics[width=16cm, bb=43 420 560 755, clip=]{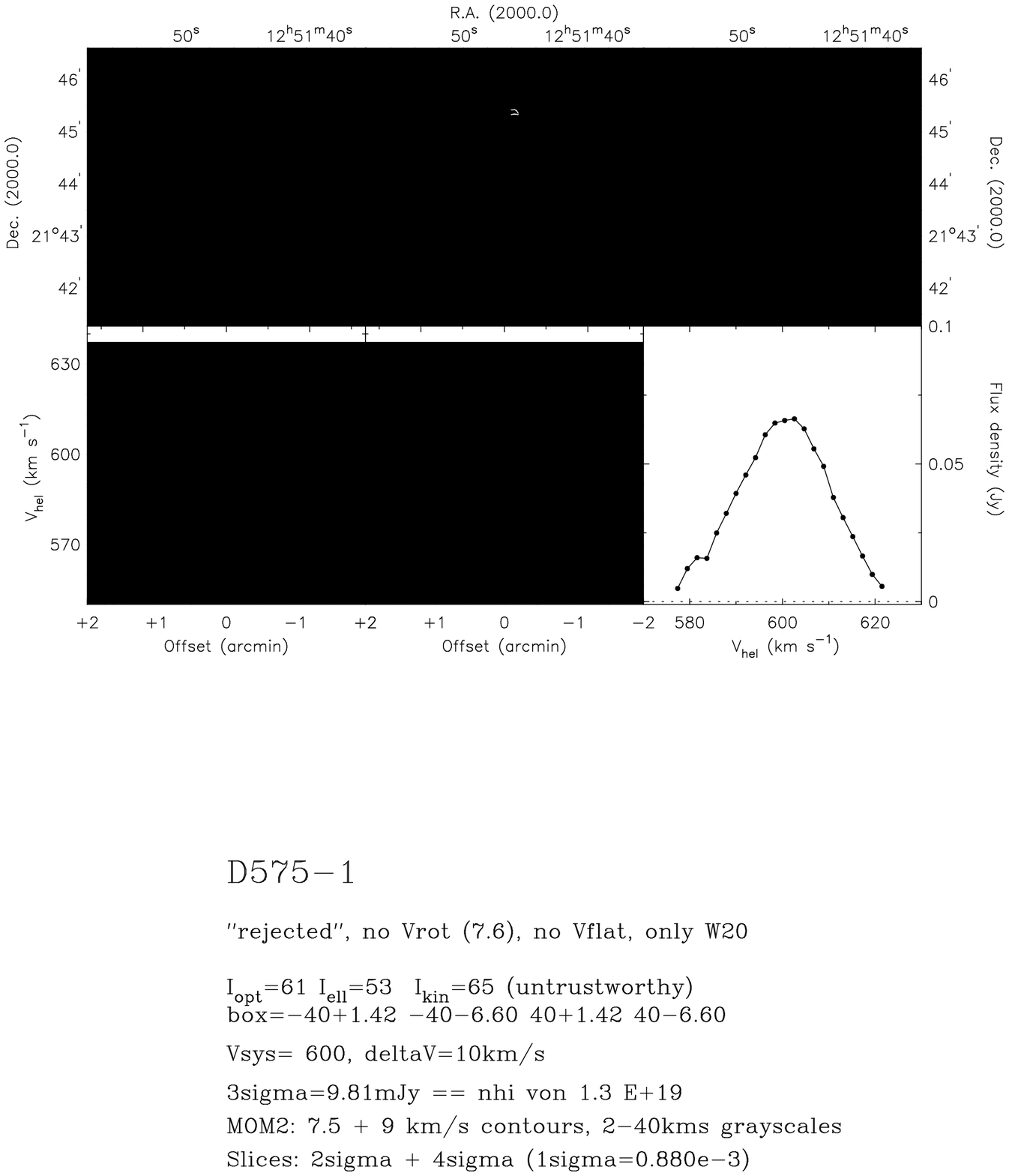}
   \caption[Summary panel for D575-1]{Summary panel for D575-1. A
     general description of the panels and levels is given at the
     beginning of the Appendix. The $3\sigma$ contour in the zeroth
     moment map corresponds to a flux of 9.81 mJy, which translates to
     a column density limit of $n_{HI}=1.3\cdot10^{19}\
     \mathrm{cm}^{-2}$. The contours in the second moment
     map are at 7.5 \kms\ and 9 \kms.  }
         \label{fig:btf:d575-1}
   \end{figure*}

\begin{figure*}
   \centering
   \includegraphics[width=16cm, bb=43 420 560 755, clip=]{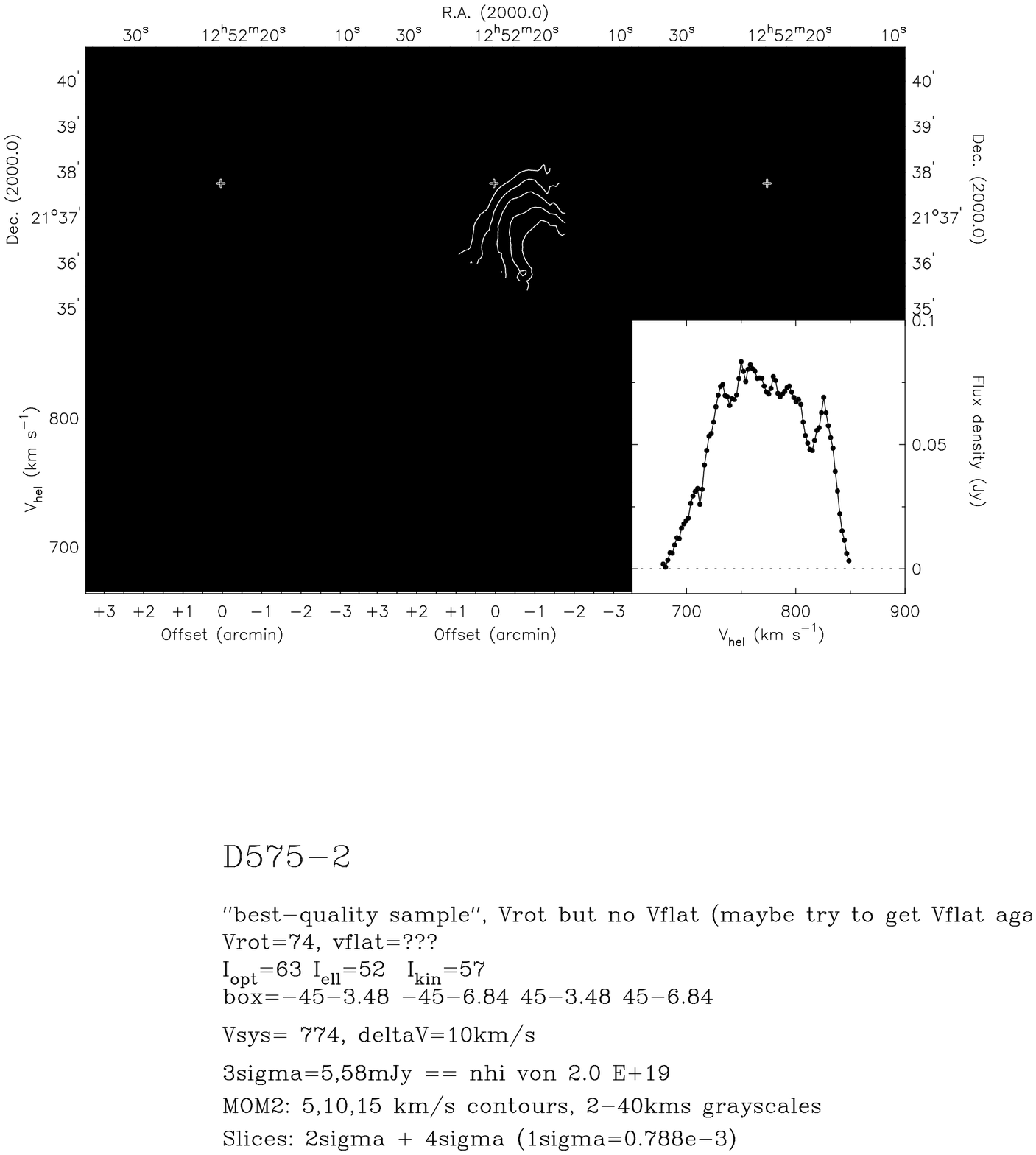}
   \caption[Summary panel for D575-2]{Summary panel for D575-2. A
     general description of the panels and levels is given at the
     beginning of the Appendix. The $3\sigma$ contour in the
     zeroth moment map corresponds to a flux of 5.58 mJy,
     which translates to a column density limit of
     $n_{HI}=2.0\cdot10^{19}\ \mathrm{cm}^{-2}$.  }
         \label{fig:btf:d575-2}
   \end{figure*}

\begin{figure*}
   \centering
   \includegraphics[width=16cm, bb=43 420 560 755, clip=]{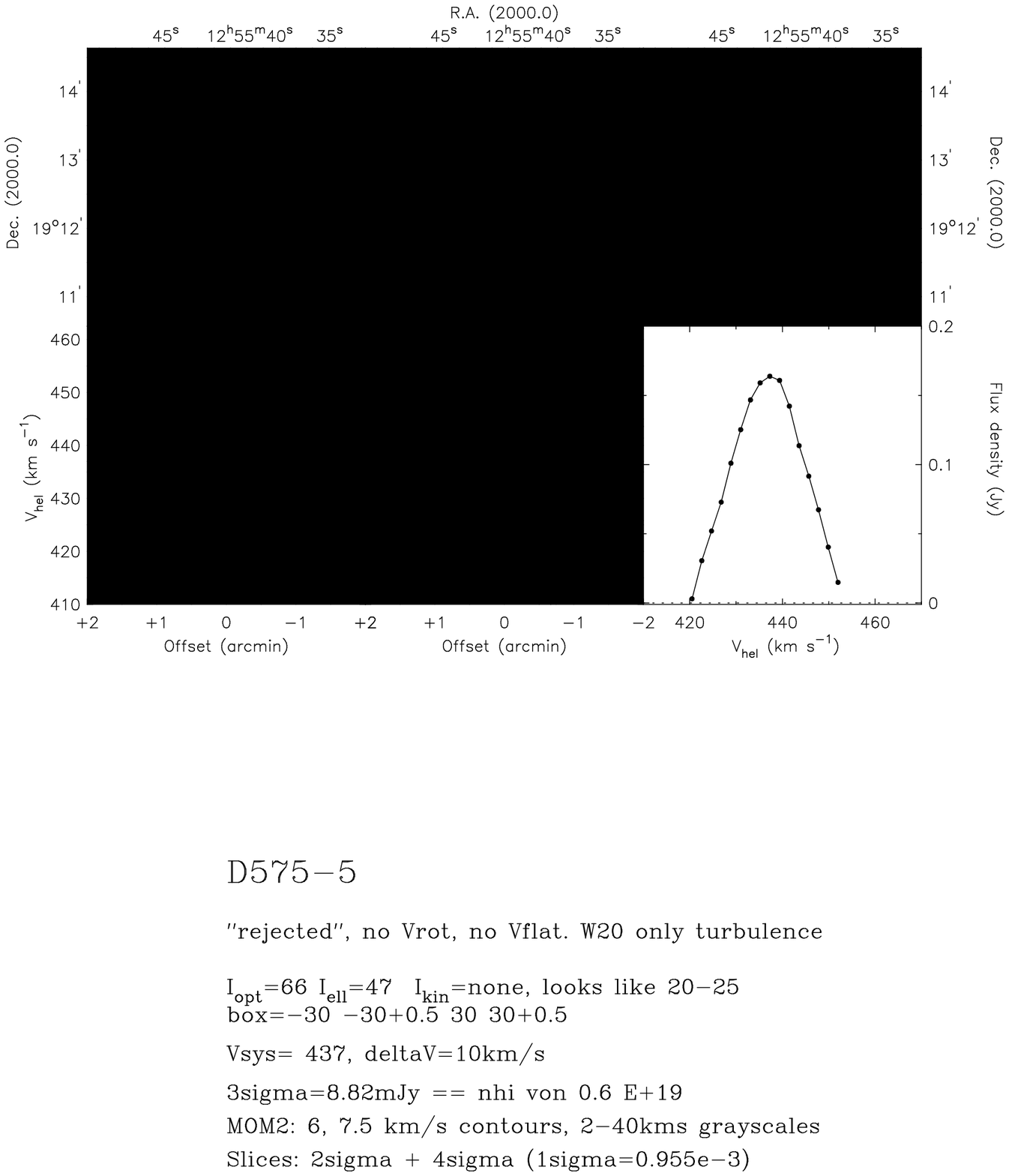}
   \caption[Summary panel for D575-5]{Summary panel for D575-5. A
     general description of the panels and levels is given at the
     beginning of the Appendix. The $3\sigma$ contour in the
     zeroth moment map corresponds to a flux of 8.82 mJy,
     which translates to a column density limit of
     $n_{HI}=0.6\cdot10^{19}\ \mathrm{cm}^{-2}$. The contours in the
     second moment map are given at 6 \kms\ and 7.5 \kms.
   }
         \label{fig:btf:d575-5}
   \end{figure*}

\begin{figure*}
   \centering
   \includegraphics[width=16cm, bb=43 420 560 755, clip=]{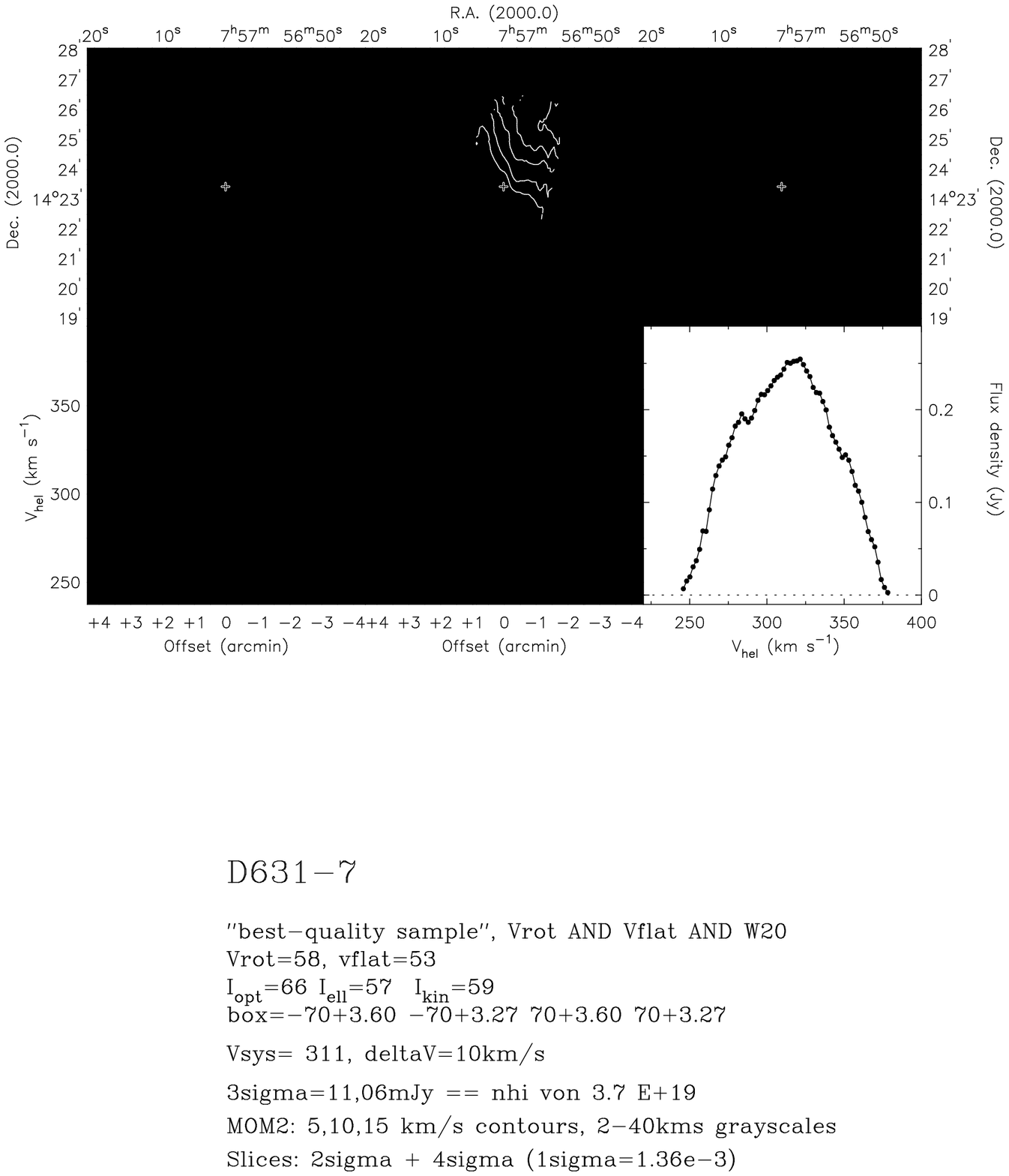}
   \caption[Summary panel for D631-7]{Summary panel for D631-7. A
     general description of the panels and levels is given at the
     beginning of the Appendix. The $3\sigma$ contour in the
     zeroth moment map corresponds to a flux of 11.06 mJy,
     which translates to a column density limit of
     $n_{HI}=3.7\cdot10^{19}\ \mathrm{cm}^{-2}$.  }
         \label{fig:btf:d631-7}
   \end{figure*}

\begin{figure*}
   \centering
   \includegraphics[width=16cm, bb=43 420 560 755, clip=]{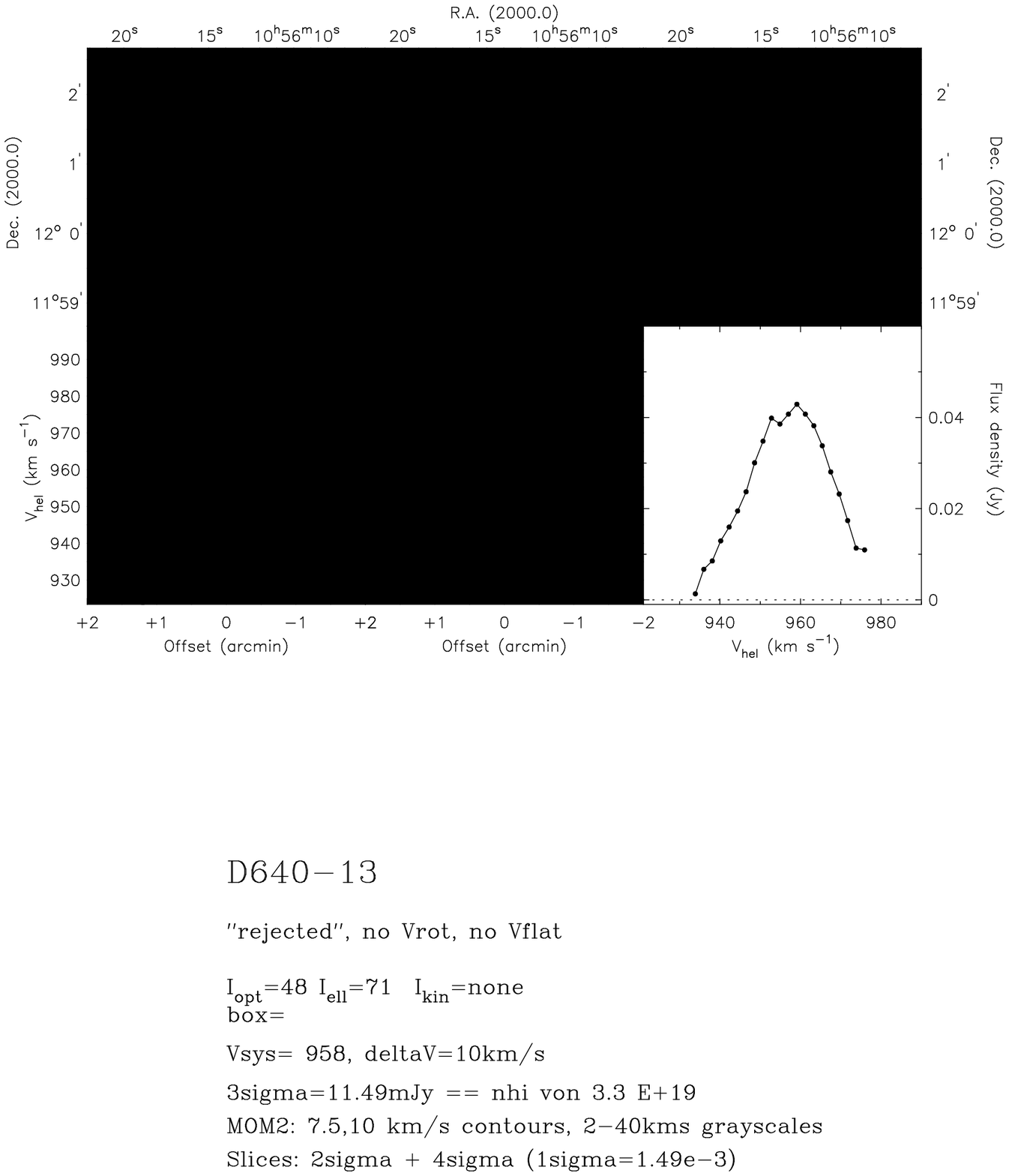}
   \caption[Summary panel for D640-13]{Summary panel for D640-13. A
     general description of the panels and levels is given at the
     beginning of the Appendix. The $3\sigma$ contour in the
     zeroth moment map corresponds to a flux of 11.49 mJy,
     which translates to a column density limit of
     $n_{HI}=3.3\cdot10^{19}\ \mathrm{cm}^{-2}$. The contours in the
     second moment map are given at 7.5 \kms\ and 10 \kms.
   }
         \label{fig:btf:d640-13}
   \end{figure*}

\begin{figure*}
   \centering
   \includegraphics[width=16cm, bb=43 420 560 755, clip=]{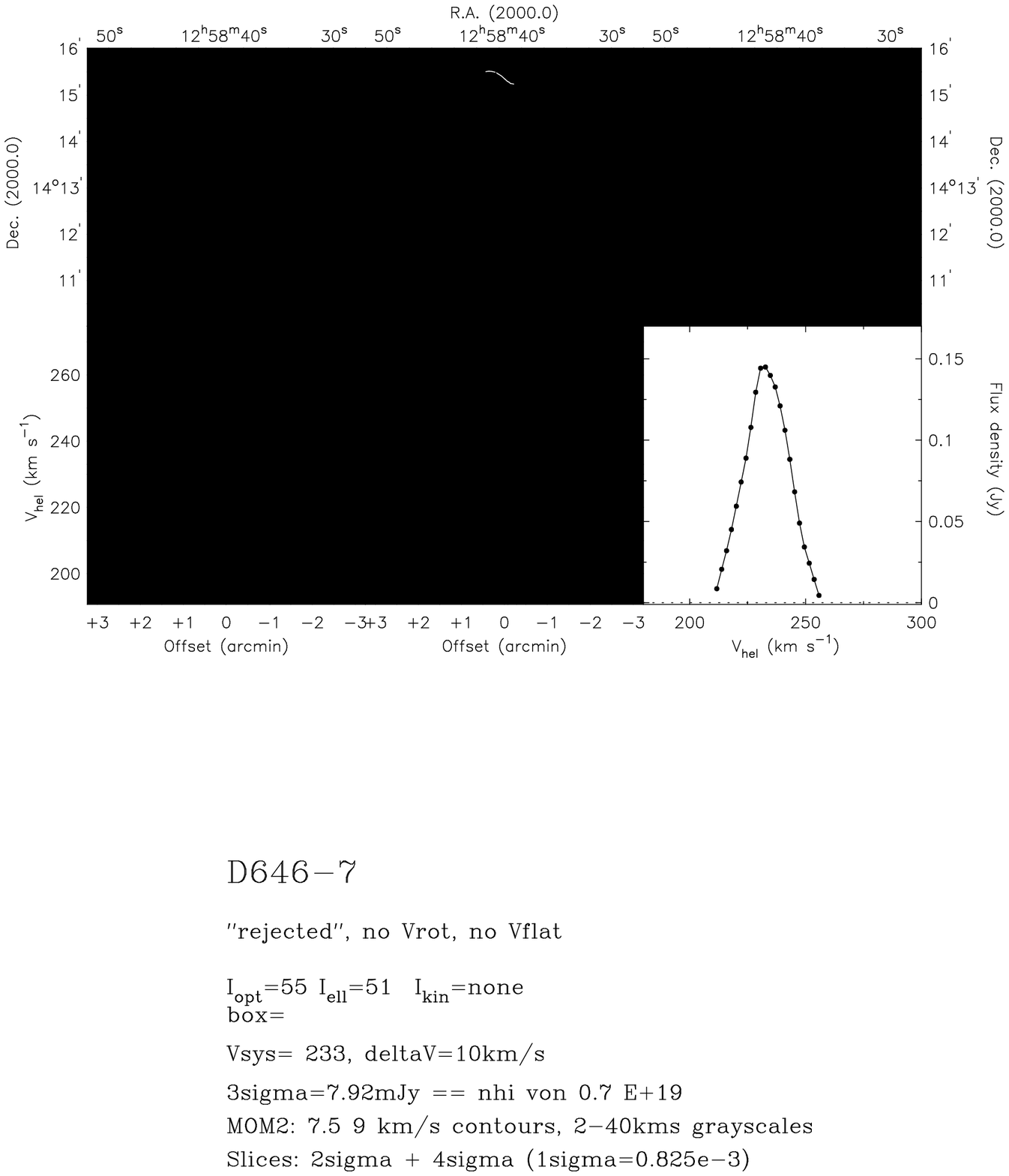}
   \caption[Summary panel for D646-7]{Summary panel for D646-7. A
     general description of the panels and levels is given at the
     beginning of the Appendix. The $3\sigma$ contour in the
     zeroth moment map corresponds to a flux of 7.92 mJy,
     which translates to a column density limit of
     $n_{HI}=0.7\cdot10^{19}\ \mathrm{cm}^{-2}$. The contours in the
     second moment map are given at 7.5 \kms\ and 9 \kms.
   }
         \label{fig:btf:d646-7}
   \end{figure*}

\begin{figure*}
   \centering
   \includegraphics[width=16cm, bb=59 165 500 690,clip=]{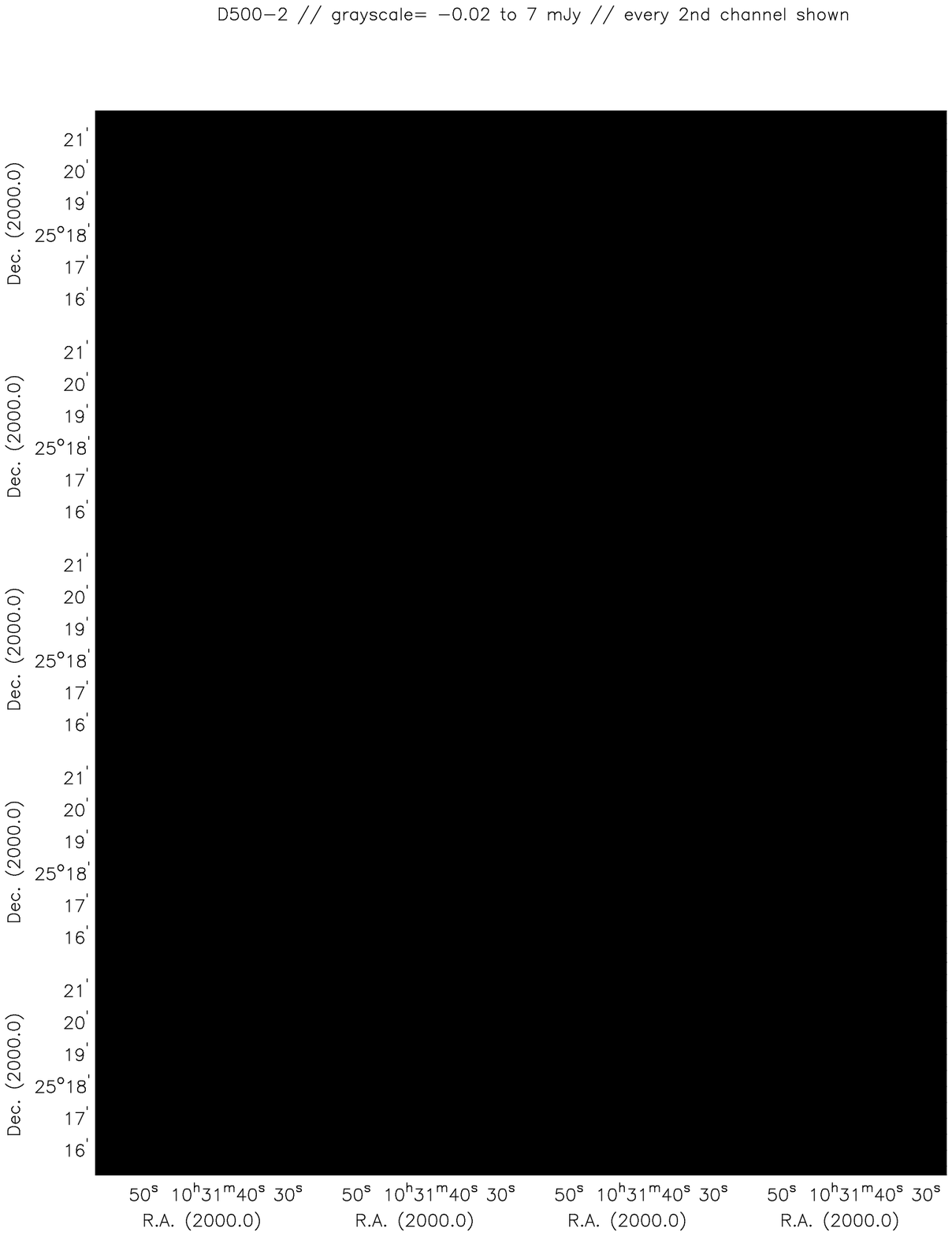}
   \caption[Channel maps of D500-2]{Channel maps of D500-2. Grayscales
     run from -0.02 to 7 mJy. Every second channel is shown.}
         \label{fig:btf:d500-2-chan}
   \end{figure*}

\begin{figure*}
   \centering
   \includegraphics[width=16cm, bb=59 165 500 690,clip=]{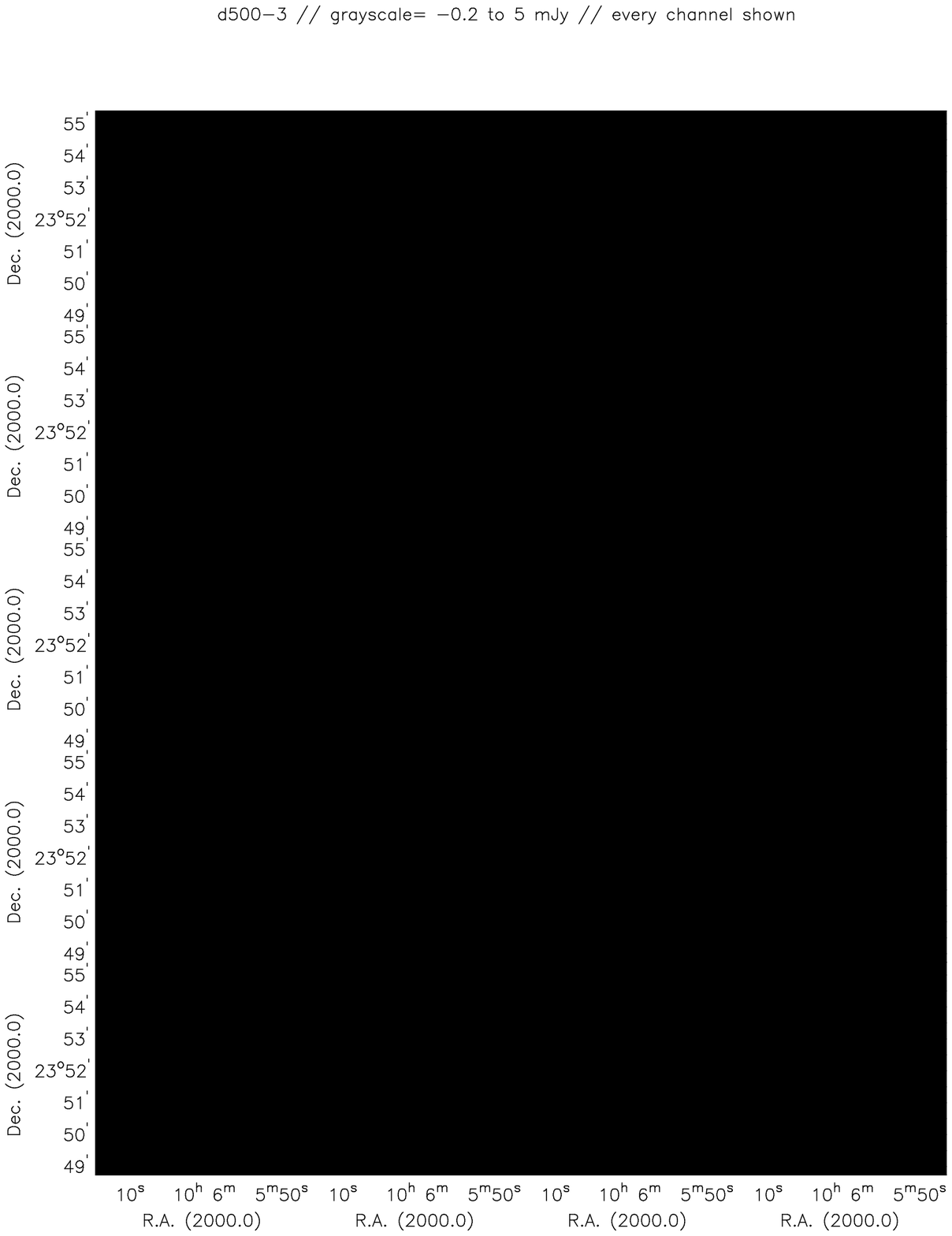}
   \caption[Channel maps of D500-3]{Channel maps of D500-3. Grayscales
     run from -0.02 to 5 mJy. Every channel is shown.}
         \label{fig:btf:d500-3-chan}
   \end{figure*}

\begin{figure*}
   \centering
   \includegraphics[width=16cm, bb=59 165 500 690,clip=]{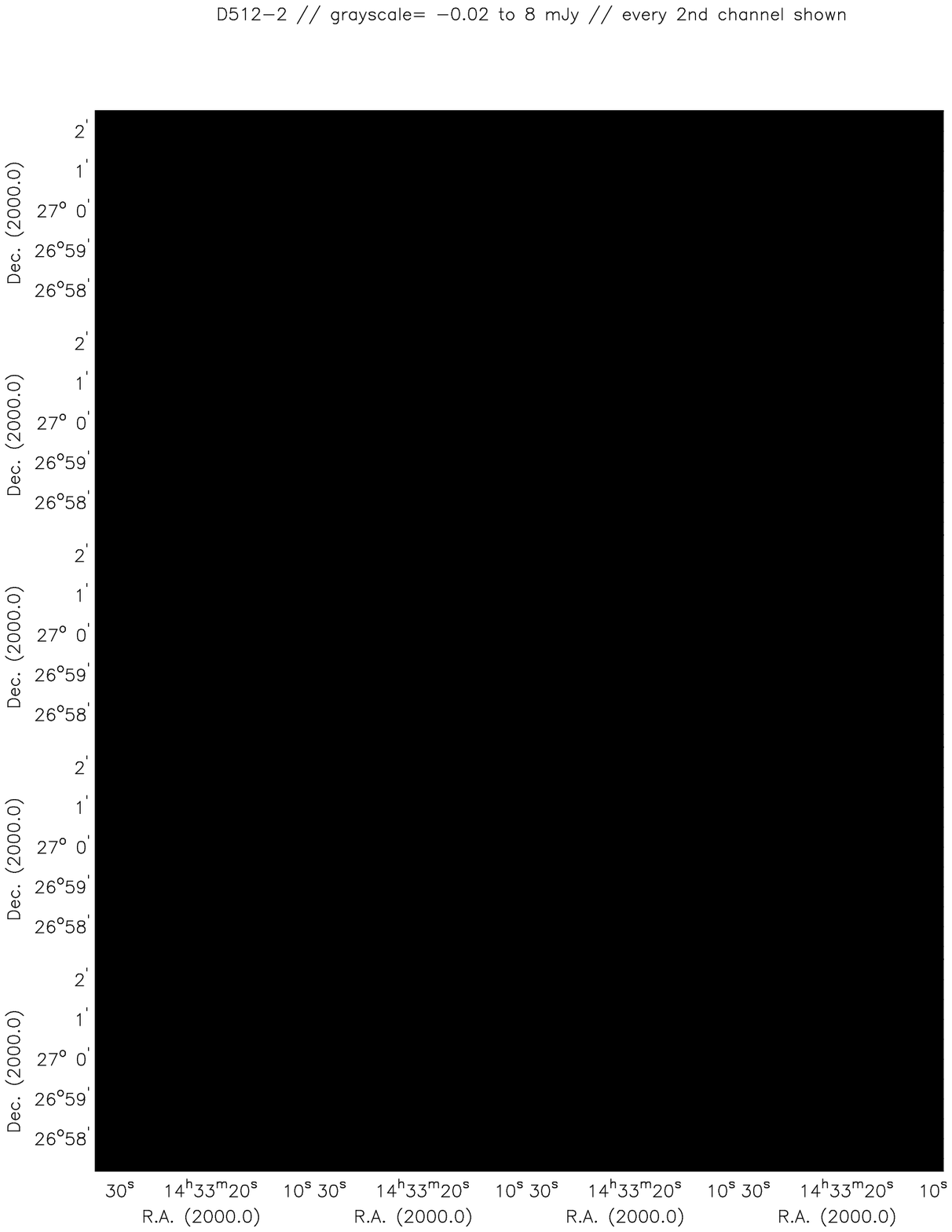}
   \caption[Channel maps of D512-2]{Channel maps of D512-2. Grayscales
     run from -0.02 to 8 mJy. Every second channel is shown.}
         \label{fig:btf:d512-2-chan}
   \end{figure*}

\begin{figure*}
   \centering
   \includegraphics[width=16cm, bb=59 165 500 690,clip=]{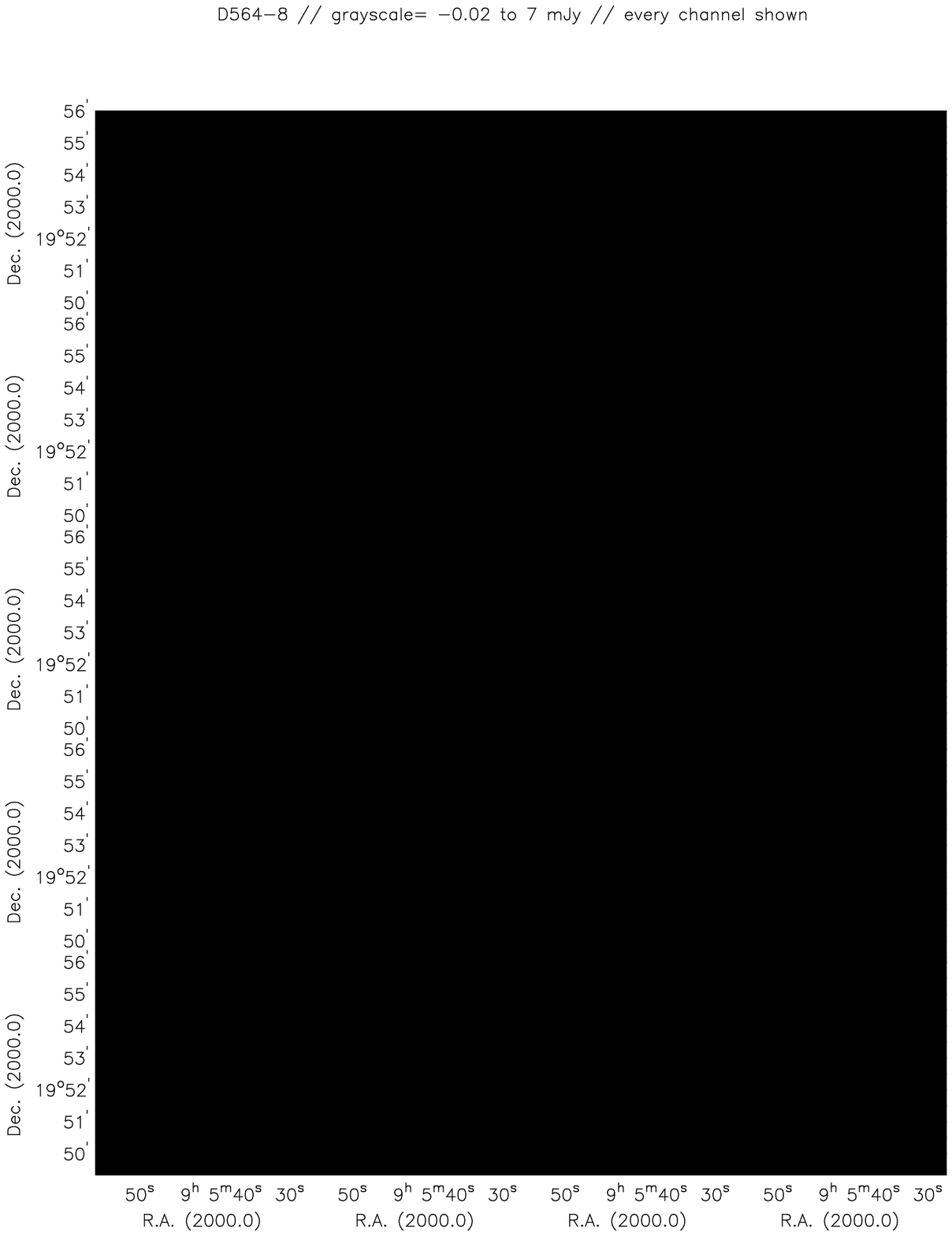}
   \caption[Channel maps of D564-8]{Channel maps of D564-8. Grayscales
     run from -0.02 to 7 mJy. Every channel is shown.}
         \label{fig:btf:d564-8-chan}
   \end{figure*}

\begin{figure*}
   \centering
   \includegraphics[width=16cm, bb=59 165 500 690,clip=]{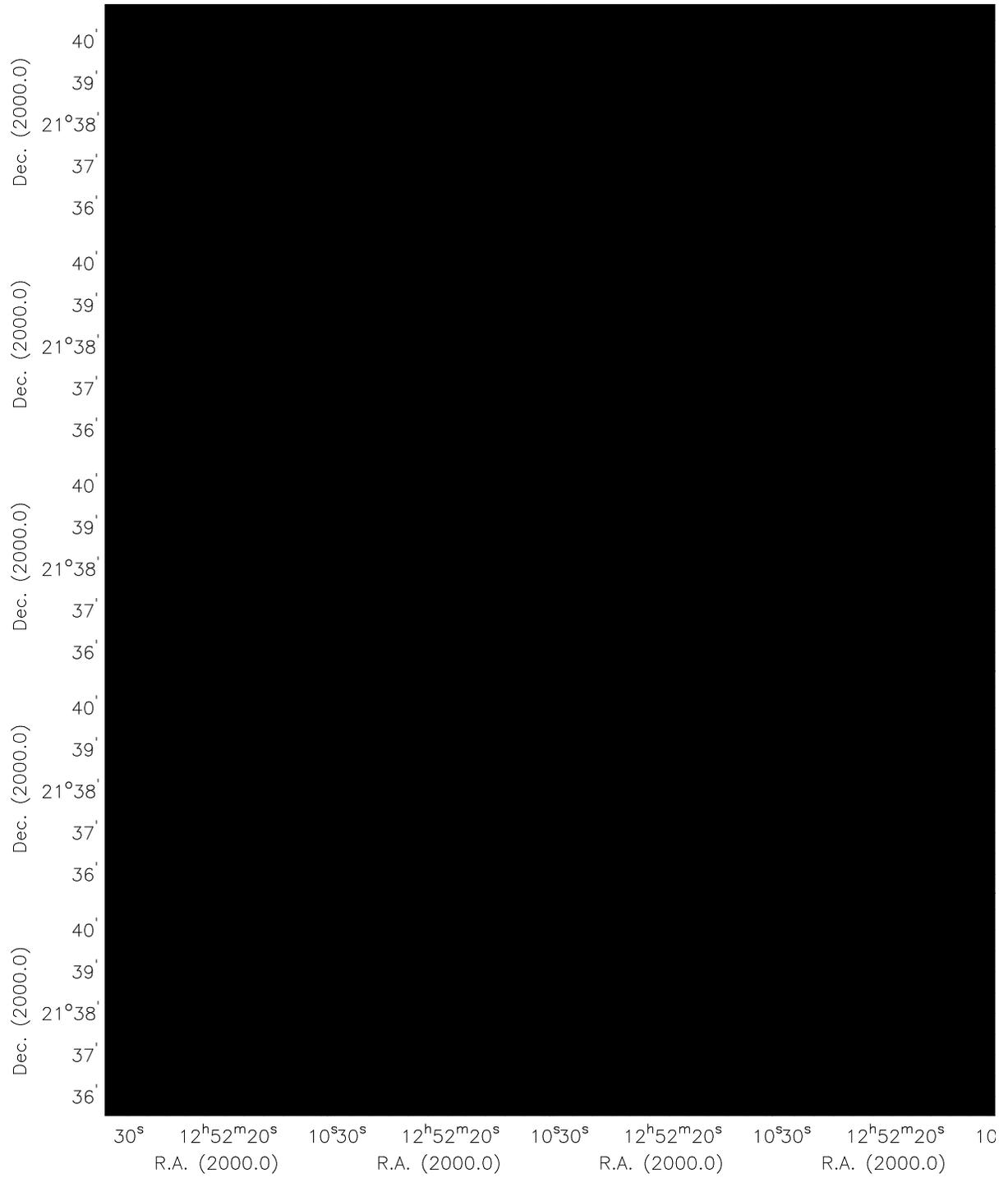}
   \caption[Channel maps of D575-2]{Channel maps of D575-2. Grayscales
     run from -0.02 to 10 mJy. Every fourth channel is shown.}
         \label{fig:btf:d575-2-chan}
   \end{figure*}

\begin{figure*}
   \centering
   \includegraphics[width=16cm, bb=59 165 500 690,clip=]{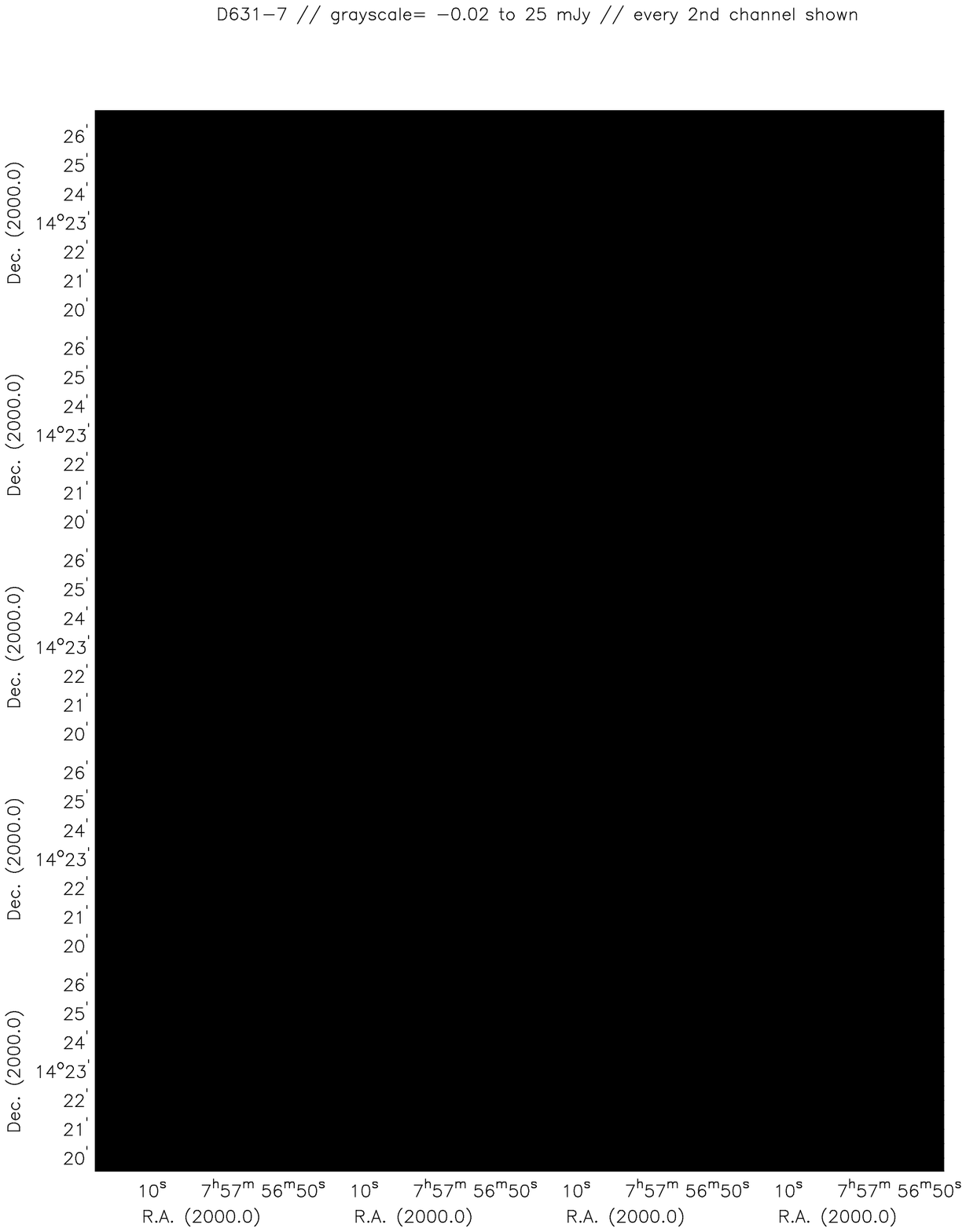}
   \caption[Channel maps of D631-7]{Channel maps of D631-7. Grayscales
     run from -0.02 to 25 mJy. Every second channel is shown.}
         \label{fig:btf:d631-7-chan}
   \end{figure*}

The Appendix contains summary panels for all galaxies of our sample
(Figs.~\ref{fig:btf:d500-2}-\ref{fig:btf:d646-7}) and channel maps for
the galaxies from the rotation curve sub-sample
(Figs.~\ref{fig:btf:d500-2-chan}-\ref{fig:btf:d631-7-chan}).

The summary panels consist of two rows with three panels each, and
contain the following maps:

{\bf Top row:} \emph{Left panel:} zeroth moment map with
grayscales. Unless mentioned otherwise, grayscales run from a column
density of $n_{HI}=1\cdot10^{19}\ \mathrm{cm}^{-2}$ (white) to
$n_{HI}=2\cdot10^{21}\ \mathrm{cm}^{-2}$ (black).  The 3$\sigma$ level
is indicated by the black contour. \emph{Middle panel:} first moment
map. The systemic velocity is indicated by the thick contour, the
contours are spaced by 10 \kms. The approaching side can be identified
by the light grayscales and black contours, and the receding side by
dark grayscales and white contours.  \emph{Right panel:} second moment
map. Grayscales run from 2 to 40 \kms. Unless mentioned otherwise, the
contours levels are given at 5, 10, and 15 \kms.  For all moment maps,
the beam size is indicated in the bottom right corner.  Additionally,
if a kinematic center has been independently derived, it is indicated
in all moment maps by a cross.

{\bf Bottom row:} \emph{Left panel:} Major axis position-velocity
diagram. The position angle of the slice indicated in the top-left
corner of the panel.  Grayscales run from 2$\sigma$ to 30$\sigma$, and
contour levels are given at $2\sigma + n \times 4\sigma$ (i.e.,
2,6,10,14,18 ... $\sigma$). The dashed line indicates the systemic
velocity.  If a \vpv\ has been derived, the resulting velocities at
either side of the rotation curve are indicated by arrows.
\emph{Middle panel:} Minor axis position-velocity diagram. Grayscales
and contours are identical to the major axis position-velocity
diagram.  \emph{Right panel:} Global \HI\ profile of the Hanning
smoothed data cube.


\end{appendix}

\end{document}